\documentstyle[times,graphics,epsfig,amsmath,amssymb,natbib,referee,psfrag]{mn2e}

 \def\S{\mathcal{S}}
\def\N{\mathcal{N}}  
\def\U{{\bf U}} \def\n{\hat{n}} \def\t{\vec{\theta}} 
\def\F{{\mathcal F}} \def\N{{\rm N}} \def\E{E} \def\snr1{ ${\rm
SNR}_1$ } \newcommand{\be}{\begin{equation}}
\newcommand{\e}{\end{equation}} \newcommand{\bear}{\begin{eqnarray}}
\newcommand{\ear}{\end{eqnarray}} 
 
  \def\apj{ApJ}
 \def\mnras{MNRAS} \def\prd{PRD}

\begin{document}

\title[Imprint of BAO on the Cross-correlation of Redshifted 21-cm
Signal and Ly-$\alpha$ Forest ]{The Imprint of the Baryon Acoustic
Oscillations (BAO) in the Cross-correlation of the Redshifted HI 21-cm
Signal and the Ly-$\alpha $ Forest}

\author[Guha Sarkar, Bharadwaj]
{ Tapomoy Guha Sarkar$^{1}$\thanks{E-mail: tapomoy@hri.res.in}
and Somnath Bharadwaj$^{2,3}$\thanks{E-mail: somnath@phy.iitkgp.ernet.in}
 \\
$^1$Harish-Chandra Research Institute, Chhatnag Road, Jhusi, Allahabad, 211019 India. \\
$^2$Department of Physics \& Meteorology, IIT, Kharagpur 721302, India.\\
$^3$ Centre for Theoretical Studies, 
IIT, Kharagpur 721302, India.\\ 
}

\maketitle \date{\today}

\begin{abstract}
The cross-correlation of the Ly-$\alpha$ forest and redshifted 21-cm
emission has recently been proposed as an observational tool for
mapping out the large-scale structures in the post-reionization era $z
\le 6$.  This has a significant advantage as the problems of
continuum subtraction and foreground removal are expected to be
considerably less severe in comparison to the respective
auto-correlation signals.  Further, the effect of discrete quasar
sampling is considerably less severe for the cross-correlation in
comparison to the Ly-$\alpha$ forest auto-correlation signal.  In this
paper we explore the possibility of using the cross-correlation signal
to detect the baryon acoustic oscillation (BAO). To this end, we have
developed a theoretical formalism to calculate the expected
cross-correlation signal and its variance.  We have used this to
predict the expected signal, and estimate the range of observational
parameters where a detection is possible. 

For the Ly-$\alpha$ forest, we have
considered BOSS and BIGBOSS which are expected have a quasar density
of $16 \ {\rm deg}^{-2}$ and $64 \ {\rm deg}^{-2}$ respectively. A  radio
interferometric array that covers the redshift range $z=2$ to $3$
using antennas of size $2 \, {\rm m} \times 2 \, {\rm m}$, which have
a $20^{\circ} \times 20^{\circ}$ field of view, is well suited for the
redshifted 21-cm observations. It is  required  to observe $25$ independent
fields of view, which corresponds to the entire angular extent of BOSS.   
We find that  it is necessary to  achieve a noise level of 
$1.1 \times 10^{-5} \ {\rm mK}^2$ and $6.25 \times 10^{-6} \ {\rm mK}^2$
per field of view  in the redshifted 21-cm observations 
to detect the angular and radial BAO respectively with BOSS. 
The corresponding figures are 
$3.3 \times 10^{-5} \ {\rm mK}^2$ and $1.7 \times 10^{-5} \ {\rm mK}^2$
for BIGBOSS. We also discuss possible observational strategies for detecting the 
BAO signal.   Four to five independent radio interferometric arrays, each containing 
$400$ antennas uniformly sampling all the baselines within $50 \, {\rm m}$ will
be able to carry out these observations in the span of a few years. 
\end{abstract}


\section{Introduction}
Neutral hydrogen (HI) in the post-reionization epoch ($z < 6$) is
known to be an important cosmological probe seen in both emission and
absorption.  Here, the redshifted 21-cm emission
\citep{hirev1, hirev3} and the transmitted QSO flux through the
Ly-$\alpha$ forest \citep{wien, Mandel} are both of utmost
observational interest. In a recent paper \citet{tgs5} have proposed
the cross-correlation of the 21-cm signal with the Lyman-$\alpha$
forest as a new probe of the post-reionization era. While it is true that the emission
and the absorption signals both originate from neutral hydrogen (HI)
at the same redshift (or epoch), these two signals, however, do not
originate from the same set of astrophysical sources. The 21-cm
emission originates from the HI housed in the Damped Lyman-$\alpha$
systems (DLAs) which are known to contain the bulk of the HI at low
redshifts \citep{proch05}. The collective emission from the individual
clouds appears as a diffuse background in low frequency radio
observations \citep{poreion2}. On the contrary, the Ly-$\alpha$ forest consists of a
large number of Ly-$\alpha$ absorption lines seen in the spectra of
distant background quasars. These absorption features arises
due to small density fluctuations in the predominantly ionized diffuse
IGM. On large scales, however, the fluctuations in the 21-cm signal
and the the Ly-$\alpha$ forest transmitted flux are both believed to
be excellent tracers of the underlying dark matter
distribution. \citet{tgs5} have proposed that the cross-correlation
between these two signals can be used to probe the power spectrum
during the post-reionization era.

The HI power spectrum can be determined separately from 
 observations  of the Ly-$\alpha$ forest \citep{pspec}
and the redshifted 21-cm emission \citep{poreion1}. 
 The detection of the individual signals, however, face
severe observational challenges. The Ly-$\alpha$ auto-correlation
power spectrum is expected to be dominated by the Poisson noise
arising from the discrete sampling of QSO lines of sight. The
cross-correlation signal has the advantage that it is not affected by
the Poisson noise which affects only its variance. Uncertainties in
fitting the QSO continuum \citep{pspec4, kim04, bolton} pose another
challenge for using Ly-$\alpha$ observations to determine the power
spectrum.  Astronomical sources like the galactic synchrotron
emission, extra-galactic point sources, etc. appear as foregrounds for
the redshifted 21-cm signal. These foregrounds, which are several
orders of magnitude larger than the signal, pose a great challenge for
detecting the post-reionization 21-cm signal \citep{fg10,fg11}. 
 The foregrounds and systematics in the Ly-$\alpha$ forest and the 
redshifted 21-cm emission  are, however expected to be
uncorrelated and we therefore expect the problem to be much less
severe for the cross-correlation signal. A detection of a 
cross-correlation signal shall, hence, conclusively ascertain its
cosmological origin.  Apart from being an independent probe of the
large scale matter distribution, the cross-correlation signal can
potentially unveil the same astrophysical and cosmological information
as the individual auto-correlations.  

Cosmological density perturbations drive acoustic waves in the
primordial baryon-photon plasma which are frozen once recombination
takes place at $ z \sim 1000$, leaving a distinct oscillatory signature
on the CMBR anisotropy power spectrum \citep{peeb70}. The sound horizon at
recombination  sets a standard ruler that maybe used to
calibrate cosmological distances.  Baryons contribute to 
 $15 \% $ of the total matter density,  and the baryon  acoustic
oscillations  are imprinted in the late time clustering of
non-relativistic matter. The signal, here, is however suppressed by a
factor $ \sim \Omega_b/\Omega_m \sim 0.1$,  unlike the  CMBR
temperature anisotropies where it is an order unity effect
\citep{komatsu}. The baryon acoustic oscillation (BAO) is  
a powerful  probe of cosmological parameters \citep{seoeisen,white}. 
This is particularly useful since the effect occurs on large scales ($\sim 150 \, 
\rm Mpc$), where the fluctuations are still in the linear regime. 
It is possible to measure the angular diameter distance and 
the Hubble parameter as functions of redshift using the  
the transverse and the longitudinal oscillations respectively. These 
provide   means for estimating cosmological
parameters and placing  stringent constraints on dark energy models.
Nonlinear effect  of gravitational clustering  tend to wipe out the BAO signal, 
and it is preferable to avoid very low redshifts where this is a potential 
problem. However, very high redshifts too are not very useful for constraining
dark energy models. 
Several authors have reported a $2$ to $3 \sigma$ detection of the BAO 
in  low redshift galaxy surveys \citep{baoeisen05, percival07, baosdss}.
The possibility of detecting the BAO signal in the Ly-$\alpha$ forest 
has  been extensively studied by \citet{cosparam1}.
 Several groups have also considered the   possibility of
detecting the BAO signal  using  redshift 21-cm emission \citep{pen2008, maowu, masui10}.  

Several QSO surveys are now being considered  with the intent of measuring the BAO  
using the Ly-$\alpha$ forest (eg. BOSS \citep{mcd1}  and 
BIGBOSS \citep{bigboss}). The possibility of a  wide field redshifted 21-cm 
survey to detect the BAO is also under serious consideration. 
In this paper, we consider  the possibility 
of studying the BAO using the cross-correlation signal.  In Section 2. of this paper
we quantify the cross-correlation between the Ly-$\alpha$ forest and the 21-cm emission,
and  present theoretical  predictions of  the expected signal. 
We have used  the multi-frequency  angular power spectrum $C_{\ell}(\Delta z)$ 
(MAPS,  \citealt{datta1})   in preference to the more commonly used three dimensional
 power spectrum $P(k)$ to quantify the cross-correlation signal. This has several
 advantages which we briefly discuss here. 
MAPS  refers to the  angular multipole $\ell$ (or  equivalently angle) and
redshift interval $\Delta z$ which are the directly relevant observational 
quantities for both  Ly-$\alpha$ surveys and redshifted 21-cm observations. This is 
particularly important if we wish  to determine 
the  angular  scales and redshift intervals  which  need to be covered in order to detect
a particular feature in the signal.  The foregrounds in the redshifted 21-cm observations
and the continuum in the Ly-$\alpha$ forest are both expected to have a smooth, slow 
variation along the frequency  axis, and this plays a crucial role in removing these 
from the respective data. It is therefore advantageous to use MAPS which retains the 
distinction between the frequency (redshift)  and angular information, unlike $P(k)$ 
 which mixes 
these up.  \citet{fg3} have analyzed GMRT observations 
using  MAPS to jointly characterize the angular and frequency dependence of the foregrounds
 at $150 \, {\rm MHz}$.  \citet{fg10} and \citet{fg11} have applied  MAPS to Analyze
$610 \, {\rm MHZ}$  GMRT observations, and have used this to characterize the 
foregrounds for the  post-reionization 21-cm signal. In fact, 
they show that it is possible  to completely remove the foregrounds 
from the  measured $C_{\ell}(\Delta \nu)$ by subtracting out polynomials in $\Delta \nu$. 
Finally, the signal could have a significant contribution from the  light cone effect, 
particularly if the observations span a large redshift interval. It is, in principle, 
possible to account for this in the MAPS, though we have not done this here. It is,
however, not possible to account for this effect in the three dimensional power spectrum   
which mixes up the information from 
different epochs through a Fourier transform along  the radial direction. 

In Section 3 of this paper we quantify the imprint of the BAO feature on the 
cross-correlation signal. Here MAPS has the  advantage that it allows us to 
separately study the radial and the transverse oscillations through the $\Delta \nu$
and the $\ell$ dependence respectively. 
In Section 4 we introduce an estimator for the MAPS of the cross-correlation signal 
and derive its statistical properties. In particular, we present
a detailed analysis of the noise for the cross-correlation estimator. In Section 5 
we present 
several observational considerations which are relevant for the cross-correlation signal,
 some   pertaining to the Ly-$\alpha$ forest and others to the redshifted 21-cm signal or 
both. Finally, in  Section 6  we discuss the detectability of the cross-correlation signal
 and the BAO features. 
We have used the cosmological parameters $(\Omega_m h^2, \Omega_b h^2,
\Omega_{\Lambda}, h, n_s, \sigma_{8})= (0.136, 0.023, 0.726, 0.71,
0.97, 0.83)$  from \citet{komatsu} throughout this paper.
\section{The Cross-correlation Signal}
We quantify the fluctuations in the transmitted flux ${\F}(\n ,z)$
along a line of sight $\n$ to a quasar through the Ly-$\alpha$ forest
using $\delta_{\F}(\n ,z)= {\F}(\n ,z)/\bar{\F}-1$. For the purpose of
this paper we are interested in large scales where it is reasonable to
adopt the fluctuating Gunn-Peterson approximation \citep{gunnpeter,
  bidav, pspec, pspec1} which relates the transmitted flux to the
matter density contrast $\delta$ as ${\F}=\exp[-A(1+\delta)^{\kappa}]$
where $A$ and $\kappa$ are two redshift dependent quantities. The
function $A$ is of order unity \citep{bolton} and depends on the mean
flux level, IGM temperature, photo-ionization rate and cosmological
parameters \citep{pspec1}, while $\kappa$ depends on the IGM
temperature density relation \citep{mac, trc}. For our analytic
treatment of the Ly-$\alpha$ signal, we assume that the measured
fluctuations $\delta_{\F}$ have been smoothed over a sufficiently large
length scale such that it is adequate to retain only the linear term
$\delta_{\F} \propto \, \delta$ \citep{bidav, pspec, pspec1,
  vielmat,saitta, slosar1}. The terms of higher order in $\delta$ are
expected to be important at small length scales which have not been
considered here.
 
We use $\delta_T(\n,z) $ to quantify the fluctuations in $T(\n,z) $
the brightness temperature of redshifted 21-cm radiation.  In the
redshift range of our interest ($z<3.5$), it is reasonable to assume
that $\delta_T(\n,z)$ traces $\delta$ with a possible bias
\citep{poreion1, poreion2}.  The bias is expected to be scale
dependent below the Jeans length-scale \citep{fang}. Fluctuations in
the ionizing background also give rise to a scale dependent bias
\citep{poreion3, poreion0}.  Moreover, this bias is found to grow
monotonically with $z$ \citep{marin}.  However, numerical simulations
\citep{bagla2, tgs2011}, indicate that it is adequate to use a constant, scale
independent bias at the large scales of our interest.

 With the above mentioned assumptions and incorporating redshift space
 distortions we may express both $\delta_{\F}$ and $\delta_{T}$ as 
\be
 \delta_{\alpha}(\n, z) = C_{\alpha} \int \ \frac{d^3
   {\bf{k}}}{(2\pi)^3} \ e^{i {\bf{k}}.\n r} [1 + \beta_{\alpha}
   \mu^2] \Delta({\bf{k}}) \,.
\label{eq:deltau}
\e 
where $\alpha={\F}$ and $T$ refer to the Ly-$\alpha$ forest and
21-cm signal respectively. Here $r$ is the comoving distance
corresponding to $z$, $\Delta({\bf{k}})$ is the matter density
contrast in Fourier space and $\mu= {\bf \hat{ k} \cdot \hat{n}}$.
 We adopt the  values $C_{\F}=-0.13$ and $\beta_{\F}= 1.58$ from
 Ly-$\alpha$ forest  simulations of \citep{mcd03}, and    
$C_T=\bar{T} \, \bar{x}_{\rm HI} \, b$ and $\beta_T=f/b$ for the 21-cm
 signal  \citep{tgs5}. These values are for  
$z=2.5$ which is the fiducial redshift in our analysis. 
  We note that there are large uncertainties in the values of all the
  four parameters $C_T, C_{\F}, \beta_T$ and $ \beta_\F$ arising from
  our poor knowledge  of the state of the diffuse IGM and the
  systems that harbour bulk of the neutral hydrogen at $z \sim 2.5$.

A QSO survey (eg. SDSS\footnote{http://www.sdss.org}), typically, covers a
large fraction of the entire sky.  In contrast, a radio
interferometric array (eg. GMRT\footnote{http://www.ncra.tifr.res.in
(check this)}
usually has a much 
smaller field of view ( $\sim 1^\circ$). Only the overlapping region
common to both these observations provides an estimate of the
cross-correlation signal.  We therefore use the limited field of view
$ L \times L $ ( $L$ in radians) of the radio telescope to estimate the
cross-correlation signal. Given this constraint,
it is a reasonable observational strategy to use several pointings of
the radio telescope to cover the entire region of the QSO survey. Each
pointing of the radio telescope provides an independent estimate of
the cross-correlation signal, which can be combined to reduce the
cosmic variance.

We have assumed that the field of view is sufficiently small ($L <<
1$) so that curvature of the sky may be  ignored.  In the flat sky
approximation the  unit vector $\n$ along  any line of sight can be
expressed as $\n = \hat{\bf{m}} + \vec{\theta} $, where $\hat{\bf{m}}$
is the line of sight to the centre of the field of view and $
\vec{\theta}$ is a two-dimensional ($2D$) vector on the plane of the
sky.  In this approximation it is convenient to decompose
$\delta_{\F}(\vec{\theta},z)$ and $\delta_T(\vec{\theta},z)$ into
Fourier modes instead of spherical harmonics.  We then have the
Fourier components 
\be \Delta_{\alpha}(\U, z ) = \int
_{-L/2}^{L/2} d^2{\vec{\theta}} \ e ^{- 2 \pi i \U \cdot
  \vec{\theta}} \ \delta_{\alpha}(\vec{\theta}, z )
\label{eq:ft}
\e 
where $\U$ is a two dimensional vector conjugate to
$\vec{\theta}$. Vector $\U$ which represents an inverse angular
scale, and also $\vec{\theta}$ are both perpendicular to the line of
sight to the centre of the field of view $\hat{\bf{m}}$.  It is useful
to visualize $\Delta_{\alpha}(\U, z )$ as Fourier components of
the signals on a plane perpendicular to the line of sight $\hat{\bf{m}}$
located at a comoving distance $r$ corresponding to the redshift
$z$. The redshift $z$, here conveys two different pieces of
information, namely the distance along the line of sight and also the
epoch where the HI signals originated.

 We introduce the multi-frequency  angular power spectrum (MAPS,
 \citealt{datta1})  
\be \langle
\ \Delta_{\alpha}(\U, z) \Delta^{*}_{\gamma}(\U ', z + \Delta z)
\ \rangle = L^2 \delta_{\U \U '} \ P_{\alpha \gamma} (\U,
 \Delta  z) 
\label{eq:crossps}
\e 
where the indices $\alpha$ and $\gamma$ have the possible values
$\alpha=T,\F$ and $\gamma=T,\F$. The multi-frequency angular power
spectrum $P_{T \F}(\U,\Delta z) \equiv P_{\F T}(\U,\Delta z)$ refers
to the cross-correlation signal, while $P_{\F \F}(\U,\Delta z)$ and
$P_{TT}(\U,\Delta z)$ refer to the respective auto-correlation
signals.  It should be noted that the multi-frequency angular power
spectra refer to the signals at two different redshifts $ z$ and $ z +
\Delta z$.  It is useful to visualize these power spectra as the
correlation of angular modes which are defined on two different planes
at comoving distances $r$ and $r + \Delta r$ corresponding to $ z$ and
$ z + \Delta z$ respectively. Our entire analysis is performed in a
narrow range of redshifts $\Delta z \ll z$. The power spectrum also
depends on $z$, however the variation with $\Delta z$ is much faster
as compared to the z dependence and the latter is not explicitly
shown. The value of $z$ for which the results have been shown is
mentioned separately wherever required.

We emphasize here that the power spectra 
$ P_{\alpha \gamma}(\U, \Delta z)$ are  directly related to 
observable quantities.  In particular, the visibilities measured in
radio interferometric observations of the redshifted 21-cm signal
\citep{poreion1, bali} are directly related to $\Delta_T( \U, z)$
defined here.  The multi-frequency angular power spectra  $ P_{\alpha
  \gamma}(\U, \Delta z)$ contain the entire three 
dimensional information through the  $\U$ and $\Delta z$ dependence.
The fluctuations in the signal in the transverse directions are
analyzed using Fourier modes $\U$ which are equivalent to the
angular multipoles in the flat sky approximation,  whereas  $\Delta z$
corresponds to a comoving separation along the radial direction.

The statistical homogeneity and isotropy of the cosmological density
fluctuations ensures that the power spectra defined above are real,
though it is not apparent from equation (\ref{eq:crossps}).  It also
follows that the power spectra are isotropic in $\U$ and depend
only on $U = |\U|$.  The assumption that both $\delta_{\F}$ and
$\delta_{T}$ are related to the matter fluctuations $\delta$ allows us
to express all the angular power spectra considered here in terms of
the three dimensional matter power spectrum $P(k)$.  Following
\citet{datta1}, 
we have 
\be 
P_{\alpha \gamma}(U, \Delta z) = \frac{1}{\pi r^2}
\int\limits_0^\infty dk_{\parallel} \ \cos(k_\parallel \Delta r) \
 F_{\alpha \gamma}(\mu) \ P(k) \,.
\label{eq:pou}
\e 
where  $\Delta r = c \Delta z /H(z)$ is the radial
comoving separation corresponding to the redshift separation $\Delta z$, $ k = \sqrt{
  k_\parallel^2 + (\frac{2\pi U}{r})^2}$, $\mu=k_{\parallel}/k$. 
and 
\be 
F_{\alpha \gamma}(\mu)=C_{\alpha} \ C_{\gamma} \ [1 + \beta_{\alpha} \mu^2]  
\ [1 + \beta_{\gamma} \mu^2]
\e

\begin{table}
\begin{center}
\begin{tabular}{ccccccccc}
$\phantom{abcde}$&  $\phantom{abcde}$ &  $\phantom{abcde}$   &$\phantom{abcde}$  &$\phantom{abcde}$
& $\phantom{abcde}$ & $\phantom{abcde}$
&$\phantom{abcde}$ & $\phantom{abcde}$ \\
\hline
\hline 
\multicolumn{4}{c}{$\phantom{abcde}$}&\multicolumn{2}{c}{$~ \quad k = 0.01~\rm Mpc^{-1}$}&\multicolumn{3}{c}{$\Delta z = 0.01$}\\
\hline
$z$& $r$ & $\nu_{21}$ & $\lambda_{Ly}$&$\ell$ & U & $\Delta r$ &$\Delta \nu$ &$\Delta v_{\parallel}$\\
$\phantom{z}$&   Mpc  &  MHz  & $\AA$  &$\phantom{\ell}$ & $\phantom{\ell}$ &  Mpc  & MHz  & $\rm Km/s$ \\
\hline 
1.5&4435  & 568 &3040 & 44 & 7 & 19 & 2.27 & 1200\\
2.5& 5944 &406 & 4256  & 59& 9.4& 12 & 1.16 & 857 \\
3.5& 6945 & 316  &5472&  69 & 11 & 8.4 & 0.70& 668 \\
\hline
\hline
\\
\end{tabular}
\label{table:tab1}
\caption{The conversions between  length scales expressed in various
  units.}
\end{center}
\end{table}

We may identify the angular mode $U$ with the angular multipole $\ell$
as $ 2 \pi U = \ell$ whereby the power spectrum $ P_a (U, \Delta z)$
can be identified with the multi-frequency angular power spectrum
(MAPS) $ {\mathcal{C}}_{\ell}^a(\Delta z) = P_a (U, \Delta z)$ under
the flat-sky approximation \citep{datta1}. The approximation is known
to be reasonably good on scales $\ell > 10$ considered in our
subsequent analysis. We shall use $\ell$ and $U$ interchangeably to
represent the inverse angular scale and similarly one may think of $
P_a (U, \Delta z)$ and $ C_{\ell}^a(\Delta z)$ to both equivalently
represent the angular power spectrum.

Quasar  surveys indicate that the redshift distribution  of the
quasars  peaks in the range   $z=2$ to  $3$ \citep{sneider}.  
For any given quasar, it is possible to estimate $\delta_{\F}$  in
only a small redshift range which is governed by the
redshift of the  quasar.   The quasar's proximity effect excludes
the region very  close to the quasar. Large redshift separations are
excluded to avoid contamination from other spectral lines. Given 
these considerations, we have chosen  $z=2.5$ as the 
fiducial redshift for our analysis.  For comparison,
we have also shown several of the
results at the neighbouring redshifts $ z = 1.5$ and $ 3.5 $.
In the subsequent discussion  we shall interchangeably use the three
dimensional 
 wave number $k$, the two dimensional Fourier mode  $U$ and the
 angular mode $\ell$ to refer to inverse angular scales on the
 sky. Similarly, it may 
 sometimes be convenient  to express the  redshift separation $\Delta
 z$ in terms of a  radial separation $\Delta r $, or a 
 frequency  separation 
$ \Delta\nu = 1420 \Delta z / {(1 + z)}^2 \rm MHz$ (for 21-cm
 observations),  or a velocity separation $\Delta v_{\parallel} = c
 \Delta z / (1 + z)$ (for Ly-$\alpha$  forest).
We have indicated  the conversion between  these various possibilities  
in Table 1. 

\subsection{Transverse Angular Power Spectrum}
\begin{figure}
\psfrag{l}[c][c][1][0]{{\bf\Large{ $\ell$}}}
\psfrag{k}[c][c][1][0]{{\bf\Large{ $k$}}}
\psfrag{P(k)}[c][c][1][0]{{\bf\Large{ $P(k)$}}}
\begin{center}
 \mbox{\epsfig{file=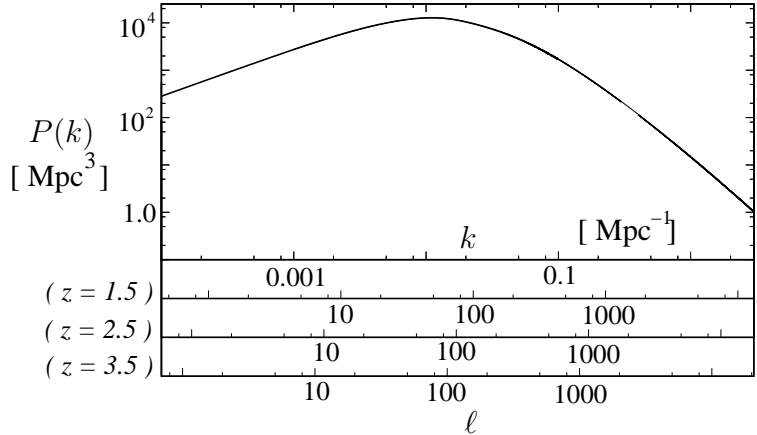,width=0.60505\textwidth,angle=0}}
\caption{ The LCDM linear power  spectrum normalised to $z=0$. The
  correspondence between $\ell$ and $k$ is shown for three different
  redshifts.  For a fixed $\ell$, all the Fourier modes to the right
  of the corresponding $k$ value contribute to $P_c(\ell,\Delta z)$.}
\label{fig:angularps1}
\end{center}
\end{figure}

\begin{figure}
\psfrag{l}[c][c][1][0]{{\bf\Large{$\ell$}}}
\psfrag{mk}[c][c][1][0]{{\bf\Large{($\rm mK$)}}}
\psfrag{mk2}[c][c][1][0]{{\bf\Large{($\rm mK^2$)}}}
\psfrag{pt}[c][c][1][0]{{\bf\Large{$P_{TT}$}}}
\psfrag{pf}[c][c][1][0]{{\bf\Large{$P_{\F T}$}}}
\begin{center}
 \mbox{\epsfig{file=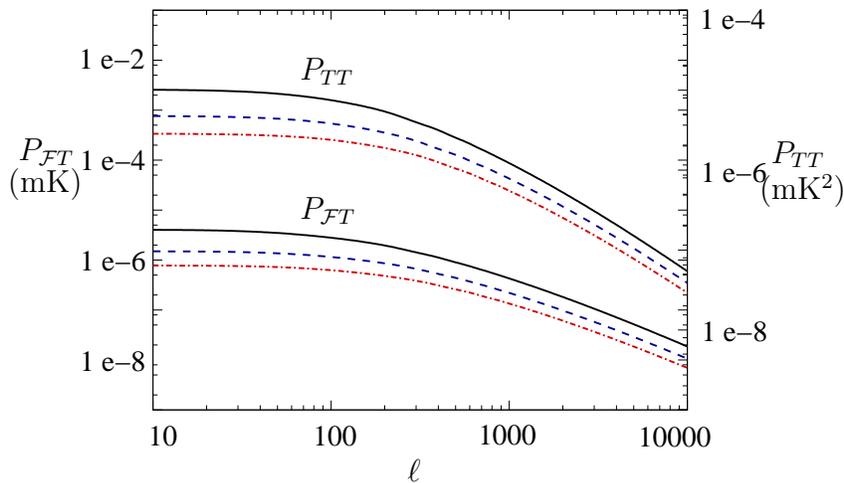,width=0.65505\textwidth,angle=0}}
\caption{ The transverse angular power spectra $P_{\F T}(\ell)$ and $P_T(\ell)$ of
  the cross-correlation and 21-cm auto-correlation signals
  respectively shown at  $ z = 1.5$ , $2.5$ and $3.5$.}
\label{fig:angularps2}
\end{center}
\end{figure}
We  first consider  $P_{\F T}(\ell) \equiv P_{\F T}  (U, \Delta z) $ for $\Delta z = 0$. The 
Ly-$\alpha$ forest and the 21-cm signals, here,   both
lie on the same plane transverse to the line of sight 
$\hat{\bf{m}}$. We refer to $P_{\F T} (\ell) $, which only contains 2D
information, as the transverse angular power
spectrum of the cross-correlation  signal. 
This has contribution
from all the three dimensional Fourier modes ${\bf{k}}$ whose projection on the 
transverse plane matches $\ell/r \equiv 2 \pi U/r $. Thus, all 3D Fourier modes  $k > \ell/r$
contribute to $P_{\F T}(\ell)$.  Figure \ref{fig:angularps1} shows the
3D matter power spectrum $P(k)$ for the LCDM model with  WMAP 5
cosmological parameters \cite{komatsu}. The correspondence between 
 $k$  and $\ell$ is shown for redshifts $z=1.5,2.5$ and $3.5$. For a
fixed $\ell$, all the Fourier modes to the right of the corresponding
$k$ value contribute to $P_{\F T}(\ell)$.  

The detectability of the cross-correlation signal is crucially
dependent on the amplitude of $P_{\F T}(\ell)$. This amplitude depends on
$C_T $ and 
$C_{\F}$, whose values are highly uncertain. Given these uncertainties
it would be very difficult to interpret the amplitude if measured,
quantitatively in terms of the physical properties of the IGM. On the
contrary it is easier to relate the shape of $P_{\F T}(\ell)$ to the matter
power spectrum and the comoving distance $r$. We elucidate this by
considering a simple toy model
for the matter power spectrum. In this model we have    $P(k) = A k$
for $k \leq k_{eq}$,  and 
 $P(k)=  Ak_{eq} (k/k_{eq})^{-3}$, for $ k >k_{eq}$. where $k_{eq}$ is
the Fourier 
 mode that enters the horizon at the epoch of  matter-radiation 
equality. Simplifying eq. (\ref{eq:pou})  by ignoring the effect
of redshift  
space distortion, $P_{\F T}(\ell)$ is the 2D projection
of the 3D matter power spectrum $P(k)$ onto the plane transverse to the line of
sight.  For large values of $\ell$ ($> k_{eq} r$) we have a power law
$ P_{\F T} (\ell) \approx  A 
C_{T}C_{\F}/{\pi \ell^2}$. The angular power spectrum flattens out
around $ \ell 
= k_{eq} r$, which corresponds to the peak in the matter power
spectrum. The position of this feature scales with redshift as $ \ell
\propto r$. At the low $\ell$ $(< k_{eq}r)$,  we have  $P_{\F T} \approx  A 
C_{T}C_{\F} /{2 \pi r^2}$ which is  independent of $\ell$. We see that there is an 
enhancement of  power at low $\ell$ (large angular scales ). This is a
generic feature  
of projection from 3D to 2D \citep{kaipea}.

Figure \ref{fig:angularps2} shows $P_{\F T}(\ell)$ for the LCDM model. 
 For comparison  we have also shown the corresponding 21-cm
 auto-correlation angular  power spectrum at the 
same probing redshifts.
The Ly-$\alpha$ auto-correlation power spectrum is  expected to be
 dominated by the Poisson  
noise arising due to the discrete QSO sampling, and this is not shown
 here. In addition  
to the fiducial redshift $2.5$ we have also shown $ z = 1.5$ and  $3.5$
with the intent of displaying  how the shape of the angular power
 spectrum changes with $z$.  
As mentioned earlier, the amplitude and also its
$z$ dependence are relatively uncertain. To estimate the amplitude we
have assumed that $C_{\F}$ and $\beta_{\F}$ do not change with $z$
while we have calculated $C_T$ and $\beta_T$ using the $z$ dependence
from \cite{tgs5}.  We see that the shape of $P_{\F T}(\ell)$ predicted for
 the LCDM 
model is in reasonable agreement with our simple toy model. At large
$\ell$ ($ >1000$) we have $P_{\F T}(\ell) \propto {\ell}^{-1.76}$ as
compared to $P_{\F T}(\ell) \propto {\ell}^{-2}$ for the toy model.  This 
discrepancy arises because the actual matter power spectrum is
shallower than $k^{-3}$ for $k>k_{eq}$.  The LCDM predictions and the
toy model are in close agreement at small $\ell$ where $P_{\F T}(\ell)$ is
nearly independent of $\ell$.  The $\ell$ value where the flattening
occurs is also found to be consistent with the predictions of the toy
model.  The behaviour of the 21-cm auto-correlation angular power
spectrum is very similar to that of the cross-correlation signal.

\subsection{Radial decorrelation function}

\begin{figure}
\begin{center}
 \mbox{\epsfig{file=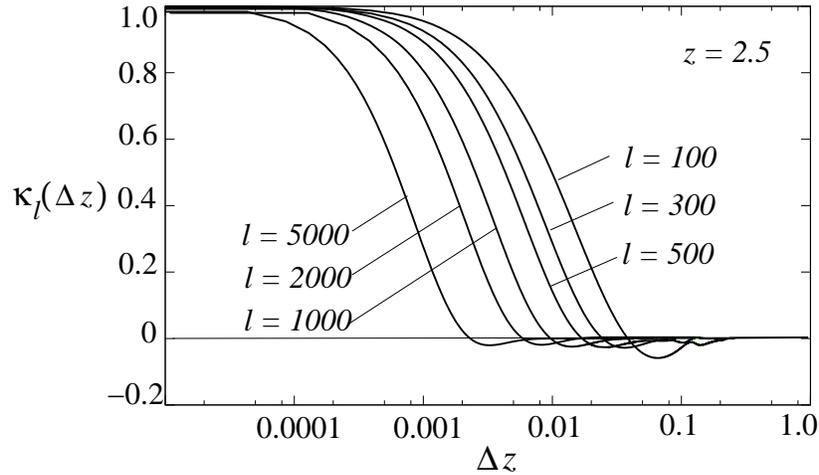,width=0.65505\textwidth,angle=0}}
\caption{The decorrelation function $\kappa_{\ell}(\Delta z)$ for 
different values of $\ell$ at our fiducial redshift $z=2.$.}
\label{fig:kappa1}
\end{center}
\end{figure}

\begin{figure}
\begin{center}
 \mbox{\epsfig{file=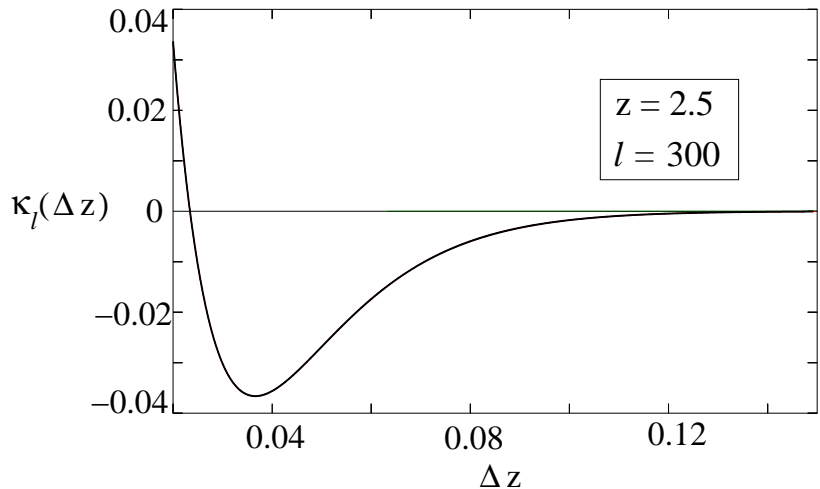,width=0.65505\textwidth,angle=0}}
\caption{The decorrelation function $\kappa_{\ell}(\Delta z)$ for 
$\ell =300$ and  $z=2.5$, showing a closeup of the
  region where the cross-correlation signal is anti-correlated.} 
\label{fig:kappa2}
\end{center}
\end{figure}

\begin{figure}
\begin{center}
 \mbox{\epsfig{file=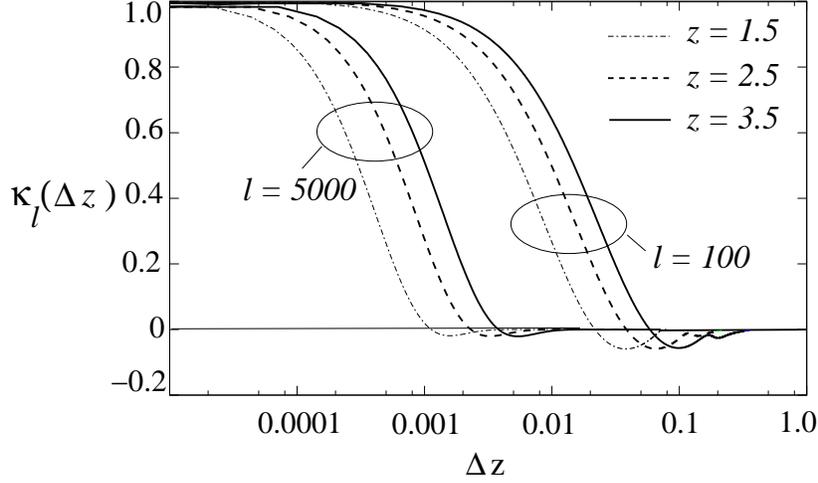,width=0.65505\textwidth,angle=0}}
\caption{For two values of $\ell$ (100 and 5000),   this   shows how
  the  decorrelation function   $\kappa_{\ell}(\Delta z)$  varies with
  $z$.}  
\label{fig:kappa3}
\end{center}
\end{figure}

Here we fix  $\ell$ (or equivalently $U$) and study $ P_{\F T} (U,\Delta z )$ as a
function of $\Delta z $. This considers the correlation between the
signals  
$\Delta_{\F}(\U,z)$ and $\Delta_{T}(\U,z+ \Delta z)$ which
refer to  the  same angular mode $\U$ but  are respectively
located  on 
two different planes  that  are separated by a comoving distance
$\Delta r$.
 The  cosine term  in eq. (\ref{eq:pou})
arises from the fact that a single  3D mode projects onto the two
different planes  with a  phase difference of $k_{\parallel}
\, \Delta r$ .  This oscillatory cosine term   ensures that  the
radial correlation  $P_{\F T}(\ell, \Delta z)$ decreases  
with increasing $\Delta z$.   Ignoring redshift space distortion, we
have $P_{\F T}(\ell, \Delta z) \propto \int_0^{\infty} \, d k_{\parallel}
\, cos(k_{\parallel} \, \Delta r) \, P(\sqrt{k^2_{\parallel} +
  (\ell/r)^2})$.  This  integral is  very close to zero if
$P(\sqrt{k^2_{\parallel} + 
  (\ell/r)^2})$  is nearly constant over the range $k_{\parallel}=0$ to 
 $k_{\parallel}=2 \pi/\Delta r$  which corresponds to  one oscillation
of the cosine term. We thus expect  
$P_{\F T}(\ell, \Delta z)$ to be maximum at $\Delta r =0$,  decrease with
increasing $\Delta r$ and remain close to zero for $\Delta r$ larger
than   $\Delta r \sim 2 \pi \, r/\ell$   \citep{poreion7}.
We quantify this behaviour using the 
decorrelation function   \citep{datta1}  
 \be 
\kappa_{\ell}(\Delta z)= \frac{P_{\F T} (\ell, \Delta z )}{P_{\F T} (\ell)}
 \e
which has value $\kappa_{\ell}(\Delta z)=1$ at $\Delta z=0$, and 
varies  in the range  $ 0 \leq |\kappa_{\ell}( \Delta 
z)| \leq 1$.  Figure \ref{fig:kappa1} shows $\kappa_{\ell}(\Delta
z)$ for different $\ell$ values at  the  fiducial redshift $z=2.5$.
We see that  $\kappa_{\ell}( \Delta z)$  falls with increasing $\Delta
z$ and  crosses zero beyond which the signal is anti-correlated 
($\kappa_{\ell}( \Delta z) < 0$).  We define the ``decorrelation
length''  $\Delta z_{0.1}$ as  the redshift separation where the
decorrelation function falls to $10 \%$ of its peak value ({\it ie.}
$\kappa_{\ell}( \Delta z_{0.1})=0.1$) .  
The signals $\Delta_{\F}(\U,z)$ and $\Delta_{T}(\U,z+
\Delta z)$  are  weakly correlated beyond the corresponding  radial
separation  $\Delta r_{0.1}$.   The decorrelation length, we find,
decreases with increasing $\ell$ {\it ie.} the signal decorrelates
faster at smaller angular scales.  This relation is well fitted by the  
relation $\Delta z_{0.1} \propto \ell^{-0.76}$ for  the entire $\ell$
and $z$ range considered here.  A similar behaviour has been reported
earlier for the 21-cm auto-correlation signal \citep{poreion7}. 

The signal crosses zero at $\Delta z > \Delta z_{0.1}$,  beyond which it
is anti-correlated.  For the  particular value  $\ell =300$,  
Figure \ref{fig:kappa2} shows a magnified view of the  region where
the signal is anti-correlated.  The behaviour is similar for other
values of $\ell$ where 
the value of $\kappa_{\ell}(\Delta z)$ varies is in the range $-0.02$
to  $-0.03$ in this region.   The  value of  $\kappa_{\ell}( \Delta z) $
 remains small,  and oscillates  around zero for very large $\Delta z$ 
(not shown in the figures). 

 The decorrelation function is sensitive to the redshift being probed.
 Figure \ref{fig:kappa3} shows  
 $\kappa_{\ell}( \Delta z) $ for three different redshifts. We find
 that though   the generic features remain    the same, 
 the decorrelation length $\Delta
z_{0.1}$ increases with   increasing redshift for all values of
$\ell$.

We note that the $\kappa_{\ell}(\Delta z)$
 for the 21-cm auto-correlation  has a behaviour very similar to
 $\kappa_{\ell}(\Delta z)$  of the cross-correlation signal discussed
 here.  In fact, the difference between these two is less than $1
 \%$. This small difference arises because of the difference in the
 values of $\beta_{\F}$ and $\beta_{T}$. 
The behaviour of the decorrelation function $\kappa_{\ell}(\Delta z)$
is sensitive to the cosmological parameters and  this can be used to 
observationally determine the cosmological parameters \citep{param3},
however we do not discuss this here. We note here that the distinct behaviour of
$\kappa_{\ell}(\Delta z)$ for the 21-cm auto-correlation allows one to distinguish
the cosmological signal from astrophysical foregrounds \citep{fg3, fg10, fg11}.

\section{The Baryon Acoustic Oscillations}
\begin{figure}
\psfrag{l}[c][c][1][0]{{\bf\Large{ $\ell$}}}
\psfrag{k}[c][c][1][0]{{\bf\Large{ $k$}}}

\psfrag{P(k)}[c][c][1][0]{{\bf\Large{ $P(k)$}}}
\psfrag{nw}[c][c][1][0]{{\bf{ $~~~\rm nw$}}}

\begin{center}
 \mbox{\epsfig{file=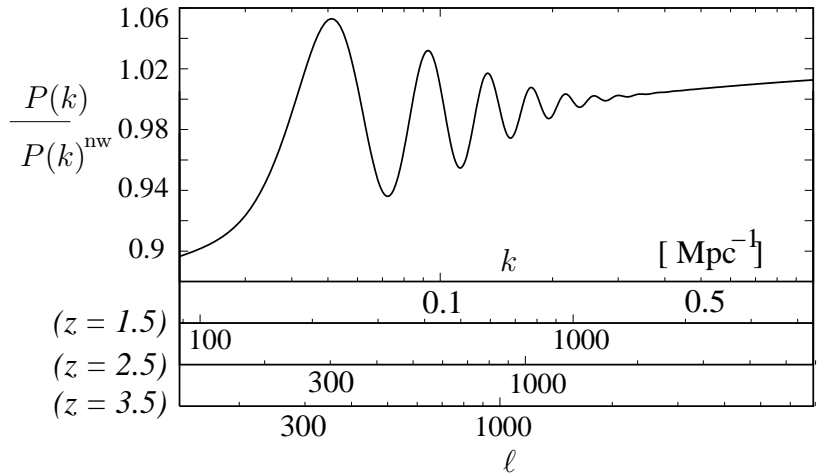,width=0.65505\textwidth,angle=0}}
\caption{The linear matter power spectrum $P(k)$, which has the BAO
  features, has been divided by $P(k)^{nw}$ the ``no wriggles'' power
  spectrum \citep{hueisen}.  The corresponding $\ell$ values have been
  shown for $z=1.5, 2.5$ and $3.5$. For a fixed $\ell$, all the Fourier modes to the right
  of the corresponding $k$ value contribute to $P_{\F T}(\ell,\Delta z)$. }
\label{fig:baopower}
\end{center}
\end{figure}

\begin{figure}
\psfrag{C}[c][c][1][0]{{\bf{ $\F T$}}}
\psfrag{P}[c][c][1][0]{{\bf\Large{ $P$}}}
\psfrag{l}[c][c][1][0]{{\bf\Large{ $\ell$}}}
\psfrag{nw}[c][c][1][0]{{\bf{ $~\rm nw$}}}

\begin{center}
 \mbox{\epsfig{file=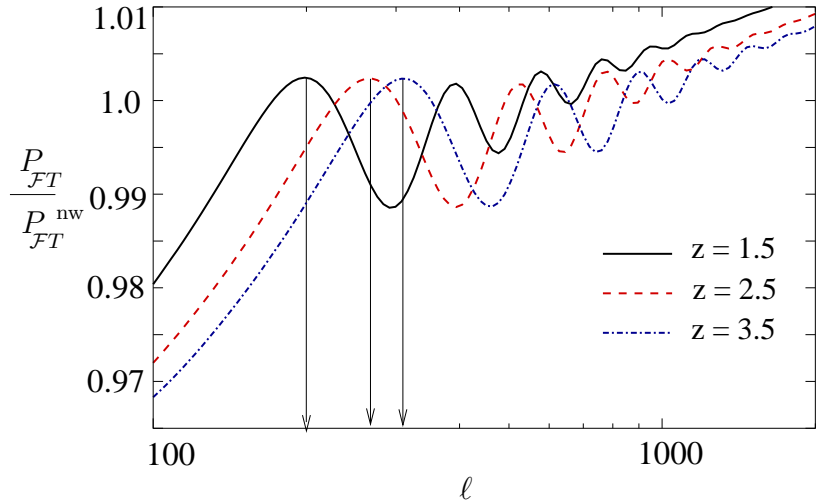,width=0.65505\textwidth,angle=0}}
\caption{This shows the imprint of the BAO  on the transverse angular
  power spectrum  $P_{\F T}(\ell)$ for the cross-correlation signal. To
  highlight the BAO we have divided $P_{\F T}(\ell)$ by  $P_{\F T}^{nw}(\ell)$
  which corresponds to $P(k)^{nw}$. This is shown for  three redshifts
  $z=1.5, 2.5$ and $3.5$. }
\label{fig:baoangular}
\end{center}
\end{figure}

The characteristic scale of the BAO is set by the sound horizon $s$ at
the epoch of recombination given by \be s = \int_a^{a_r}
\frac{c_s(a)}{a^2 H(a)} da \e where $a_r$ is the scale factor at the
epoch of recombination and $c_s$ is the sound speed given by $c_s(a) =
c/\sqrt{3(1 + 3 \rho_b/4 \rho_\gamma)}$, where $\rho_b$ and
$\rho_\gamma$ denotes the photon and baryonic densities respectively.
The comoving length-scale $s$ defines a transverse angular scale $
\theta_s = s [(1+z) D_A(z)]^{-1}$ and a radial redshift interval
$\Delta z_s =s H (z)/c$, where $D_A(z)$ and $H(z)$ are the angular
diameter distance and Hubble parameter respectively.  The comoving
length-scale $s= 143 \, \rm Mpc$ corresponds to $\theta_s=
1.38^{\circ}$ and $\Delta z_s = 0.07$ at $z = 2.5$.  Measurement of
$\theta_s$ and $\Delta z_s$ separately, allows the independent
determination of $ D_A(z)$ and $H(z)$. The CMBR angular power spectra
maps the density fluctuations on the plane of the sky at $ z \sim
1000$. Hence, the imprint of BAO in the form of acoustic peaks in the
CMBR anisotropy angular power spectrum is only in the transverse
direction and constrains $ \theta_s $.  Low redshift galaxy surveys
contain both the transverse and the radial BAO peaks.  However, the
difficulties in probing large radial distances leads to small survey
depths. Moreover, a large shot noise contribution degrades the SNR
making it very difficult to independently measure $D_A(z)$ and
$H(z)$. Typically, the combination ${[(1+z)^2 D_A^2(z)c
    z/H(z)]}^{1/3}$ is measured instead in galaxy redshift surveys
\citep{eisen05, percival07}.  The multi-frequency angular power spectrum
discussed here allows us, in principle, to measure the BAO imprint in
the both transverse and the radial directions.  Large bandwidth radio
observations covering a large portion of the sky, along with high
density quasar surveys with spectra measured at high SNR would allow
a detection of the BAO feature in both radial and transverse
directions.  The different sensitivities of the measured quantities
$D_A(z)$ and $ H(z)$ on the cosmological parameters would ensure
breaking of degeneracies in the parameter space.  For example $D_A(z)$
shall constrain curvature more efficiently than $H(z)$.  Further,
independent measurement of $D_A(z)$ and $H(z)$, would allow an
internal cross check thereby making parameter estimation self
consistent.

The BAO manifests itself as a series of oscillations in 
the linear matter power spectrum  \citep{hueisen}.  We focus on the 
first BAO peak which has the largest amplitude. 
 This  is a  $\sim 10\%$ feature in $P(k)$ at $k
 \approx 0.045$ (Figure \ref{fig:baopower}) which  corresponds to
$\ell \approx 270$ at $z=2.5$.  We also see that the first peak  will
 shift to lower  $\ell$ at smaller redshifts.

Figure \ref{fig:baoangular} shows the BAO feature in the transverse
angular power spectrum  of the cross-correlation
signal $P_{\F T}(\ell)$. The BAO, here, is seen projected onto a 
plane.  The BAO   appears as a series of oscillations in $P_{\F T}(\ell)$, the
  positions of  the peaks being  consistent with $\ell \sim  k/r$.
  The amplitude of the first oscillation in $P_{\F T}(\ell)$  is around  $1
  \%$, in   contrast to  the $\sim 10 \%$ feature seen in $P(k)$.   
This reduction in amplitude arises due to  the projection to a plane
whereby several 3D Fourier  modes  which do not have the 
  BAO feature also contribute to the $\ell$ where the first BAO peak
  is seen.   At $z=2.5$ the
  first peak occurs at $\ell \sim 270$ and it has a full width of
  $\Delta \ell \sim 200$.  The position  $\ell$ and  width
  $\Delta \ell$ 
  of the peak both scale  as $r$ if the redshift is changed. It is, in
  principle, 
  possible to determine $D_A(z)$ by measuring the $\ell$ position of
  the first BAO peak.

\begin{figure}
\begin{center}
 \mbox{\epsfig{file=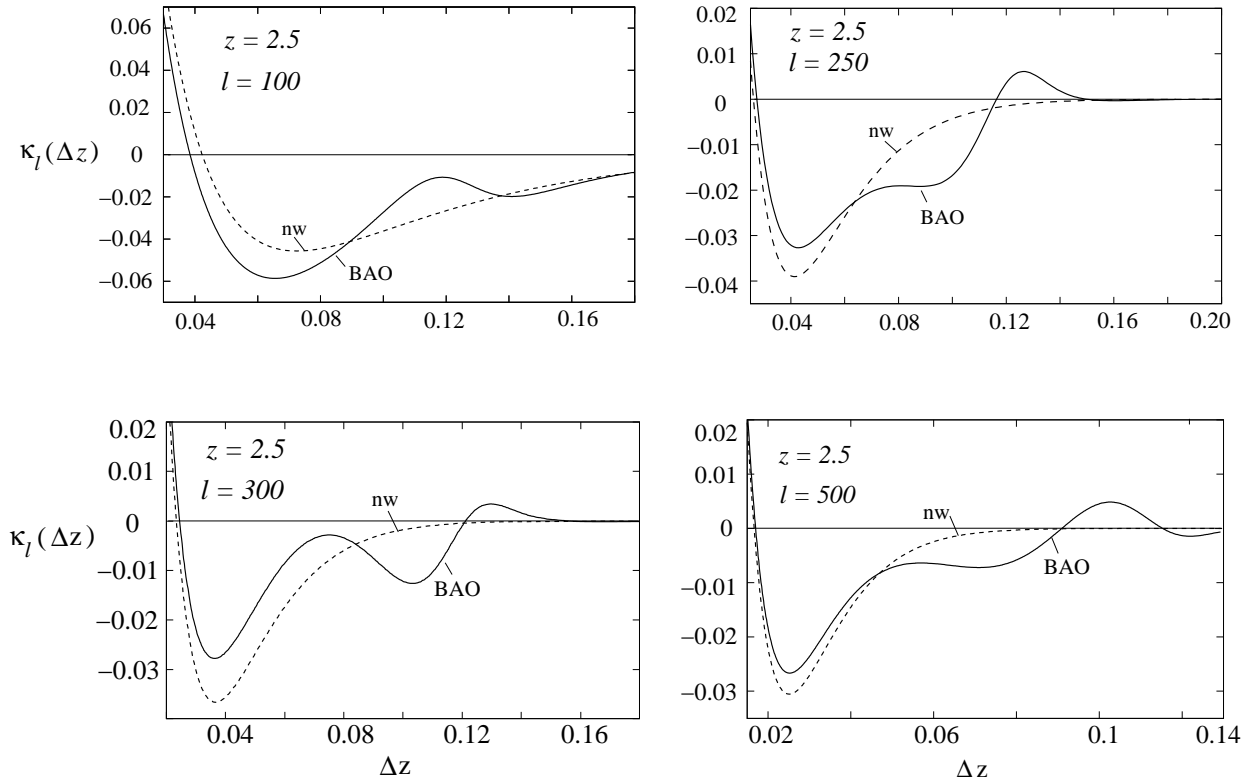,width=1.00\textwidth,angle=0}}
\caption{The solid curves  shows the imprint of BAO on the radial
  decorrelation function $\kappa_{\ell}(\Delta z)$  
for a few representative $\ell$s. The no-wiggles model (dashed curves)
  is also shown for comparison.}
\label{fig:baokappa}
\end{center}
\end{figure}

We next consider the imprint of the BAO on the cross-correlation
signal along the radial direction. This is quantified, as before,
through the 
radial decorrelation function $\kappa_{\ell}(\Delta z)$. 
We expect the first BAO peak to have an imprint  on
$\kappa_{\ell}(\Delta z)$ at only the angular modes  $ \ell$ which are
less than $ k_1 r$, where $k_1$   refers to the position of the
first peak  in the 3D matter  power spectrum.
Thus,  for $z = 2.5$  we do not expect the first peak to have any
impact on $\kappa_{\ell}(\Delta z)$ at  $\ell > 500$. We expect the
first BAO peak to have an impact at all angular modes $\ell \le 500$.
The relative contribution is expected to be maximum around  
$\ell \sim 270$, and then fall off at $ \ell < 100$ where it will be
diluted by  the $k$ modes which are  smaller than 
$k_1$ and hence do not contain the BAO signal. 
Figure \ref{fig:baokappa}  shows the radial decorrelation function
$\kappa_{\ell}(\Delta z)$  for four 
different values of $\ell \, (= 100, 250, 300,500)$.  For comparison,
we also show $\kappa_{\ell}(\Delta z)$ calculated using $P^{nw}(k)$
which does not have the BAO features. 

We recollect that $\kappa_{\ell}(\Delta z)$ is maximum at $\Delta z =
0$ where $\kappa_{\ell}(\Delta z)=1$. In the absence of the BAO
features the value of $\kappa_{\ell}(\Delta z)$  falls rapidly with
increasing $\Delta z$. This is followed by a $\Delta z$ range where 
 $\kappa_{\ell}(\Delta z)$ is negative, beyond which it 
 oscillates around $0$.   The BAO, we find, has little  impact on the
 $\Delta z$ range near $\Delta z=0$ where $\kappa_{\ell}(\Delta z)$ is
 positive. The BAO features are found to have a very significant
 effect at large $\Delta z$, typically in the range $\Delta z= 0.04$
  to  $0.16$, where $\kappa_{\ell}(\Delta z)$
 is mainly negative. The corresponding 21-cm frequency and radial velocity
 intervals are $\Delta \nu = 4.64 \, - \, 13.9 \, {\rm MHz}$ and
 $\Delta v_{\parallel} = 3.5 \times  10^3 \, - \, 10.3 \times 10^3 \,
 {\rm km/s}$ respectively. 
 The imprint of the BAO  appears as a  ringing
 feature  around the smooth `no wriggles' $\kappa_{\ell}(\Delta z)$.
 The ringing feature due to the BAO is quite distinct from the 
slow oscillation of $\kappa_{\ell}(\Delta z)$ which is also 
present in the  no-wiggles model.  The value of
$\kappa_{\ell}(\Delta z)$ is quite small ($\sim -0.01$) in the $\Delta
z$ range where the BAO features occur.  The deviation due to the 
BAO, however, could be as large as $40 \%$ to $\sim 100 \%$ 
 relative to the no-wiggles model. 

The position of the BAO feature shifts to smaller  $\Delta z$ values
if  $\ell$ is increased.  For a fixed
$\ell$, we use  $\Delta z_{\rm BAO}$ to denote  the
position  corresponding to the maximum deviation from the no-wiggles
model.      The value 
$\Delta z_s = H (z) s/c$, introduced earlier, corresponds to 
$\Delta z_{\rm BAO}$ at $\ell=0$.  It is not possible to directly
determine $\Delta z_s$  from $\kappa_{\ell}(\Delta z)$. It is
However, it is in principle possible to determine $H(z)$ by
measuring $\Delta z_{\rm   BAO}$ at different $\ell$ values.

\section{The Cross-correlation Estimator}
 In this section we  construct an estimator for $P_{\F T}(U,\Delta
 z)$,  
and consider the statistical properties of this estimator. First, we
assume that both the Ly-$\alpha$  forest and the 21-cm observations
are pixelized along  the $z$ axis into pixels or 
channels of width $\Delta z_c$ such that both $\delta_{\F}(\t,z)$ and 
$\delta_{T}(\t,z)$ are measured  only at discrete redshifts $z_n=z_0 +
n \Delta z_c$ with $n=1,2,...,N_c$. Here $z_0$ is a reference
redshift, $N_c$ is the 
total number of channels and $N_c \, \Delta z_c=B$ is 
the total redshift interval or the bandwidth spanned by the
observations.  

 Considering
first the Ly-$\alpha$ forest, we have, till now,  considered
$\delta_{\F}(\t,z_n)$ as a continuous field defined at all points on
the sky. In reality, it is possible to measure this only  along a few,
discrete lines of sight where there are background quasars.  We 
account for  this by defining $ \delta_{{\F} o}(\vec{\theta},n) $. 
the observed fluctuation in the transmitted Ly-$\alpha$ flux, as 
 \be \delta_{{\F} o}(\vec{\theta},n) =
\rho(\vec{\theta}) \, [ \, \delta_{\F}(\vec{\theta}, z_n) + \delta_{{\F}
  \N}(\vec{\theta},n) ] 
\label{eq:e1}
\e 
where $\delta_{{\F}  \N}(\vec{\theta}, z)$ is the contribution from
the pixel noise  in the quasar spectra and 
 \be \rho(\vec{\theta}) =\frac {\sum_a
  w_a \ \delta_D^{2}( \vec{\theta} - \vec{\theta}_a) }{\sum_a w_a} 
\label{eq:e2}
\e
is the quasar sampling function. Here $a=1,2,...,N$ refers to the 
different quasars in the   $L \times L$ field of view,
 $\vec{\theta}_a$ and $w_a$ respectively refer  to the angular
positions and weights of the individual quasars. We have the freedom
of adjusting the weights to suit our convenience. 
It is possible to
change  the relative contribution from the individual quasars by
adjusting the weights $w_a$.

The quasar sampling function $\rho(\vec{\theta})$ is zero everywhere
except  the angular  position of the different quasars.  It is
sometimes convenient to express the noise contribution in
eq. (\ref{eq:e1}) as
\be 
\rho(\vec{\theta}) \,  \delta_{{\F}
  \N}(\vec{\theta},n) = \frac{ 
\sum_a  w_a \ \delta_D^{2}( \vec{\theta} - \vec{\theta}_a) \
\delta_{{\F}  \N}(\vec{\theta}_a,n)}{\sum_a  w_a }
\label{eq:e3}
\e
where $\delta_{{\F}  \N}(\vec{\theta}_a,n)$ refers to the pixel noise
contribution for the different quasars. 
The faint quasars typically have
noisy spectra in comparison to the bright ones.  We can  take this
into account and choose the  weights $w_a$ so as to increase the
contribution from  the bright quasars relative to the faint ones,  
thereby maximizing  the SNR for  the signal estimator.   
For the present analysis we have
made the simplifying assumption that the magnitude of $\delta_{{\F}
  \N}(\vec{\theta}_a,n)$ is the same across all the quasars
irrespective of the quasar flux. We have modelled   $\delta_{{\F}
  \N}(\vec{\theta}_a,n)$ as Gaussian random variables with the noise
in the different pixels being uncorrelated {\it ie.}
\be \langle \delta_{{\F}  \N}(\vec{\theta}_a,n) \ \delta_{{\F}
  \N}(\vec{\theta}_b,m)  \rangle = \delta_{a,b} \delta_{n,m}
\sigma^2_{\F\N} 
\label{eq:e4}
\e
where $\sigma^2_{\F\N}$ is the variance of the pixel noise
contribution. It is appropriate to use uniform weights $w_a=1$ 
in this situation. 
Such an assumption is  justified in the  situation where  there exist 
high SNR measurements of the  transmitted  flux for all 
the quasars.

Considering next the sampling function $\rho(\vec{\theta})$, we assume
that the  quasars are randomly distributed with no correlation amongst
their angular position, and the positions  also being unrelated to
$\delta_{\F}$ and $\delta_{T}$.  In reality, the quasars do exhibit
clustering \citep{myers},  however the contribution from the Poisson
fluctuation is considerably more significant here and it is quite
justified to ignore the effect of quasar clustering for the present
purpose. The Fourier transform of $\rho(\vec{\theta})$ then has the
properties that 
\be 
\langle \tilde{\rho}(\U) \rangle = \delta _{\U,0}
\label{eq:e5}
\e
and 
\be 
\langle \tilde{\rho}(\U_1) \tilde{\rho}(\U_2)
\rangle = \frac{1}{N} \, \delta _{\U_1,\U_2} + \left( 1-\frac{1}{N} \right) 
\, \delta _{\U_1,0} \delta _{\U_2,0}
\label{eq:e6}
\
\e
which we shall use later. In the subsequent analysis, we also assume
that $N \gg1 $ whereby $(1-1/N) \approx 1$.

We use  $ {\Delta}_{{\F} o}(\U,z)$ to denote  the Fourier
of $\delta_{{\F} o}(\vec{\theta}, z)$.  Using eq. (\ref{eq:e1}) we
have 
 \be 
{\Delta}_{{\F}  o}(\U,z) = \tilde{\rho}(\U) \otimes
[ {\Delta_{\F}}(\U, z) + \Delta_{N\F}(\U, z)]
\label{eq:e7}
\e where  $ \otimes $ denotes a convolution defined as
\be 
\tilde{\rho}(\U) \otimes {\Delta_{\F}}(\U, z) =
\frac{1}{L^2} \sum_{\vec{U'}} \ \tilde{\rho}(\U - \vec{U'})
     {\Delta_{\F}}(\vec{U'}, z)
\label{eq:e8}
 \e 

Using eqs. (\ref{eq:crossps}), (\ref{eq:e3}),
(\ref{eq:e4}), (\ref{eq:e5}), (\ref{eq:e6})  and 
(\ref{eq:e7}) we calculate the following statistical properties of
$\Delta_{{\F} 0}$, 
\be 
 \langle \Delta_{{\F} o}(\U,n) \rangle =0
\label{eq:e9}
\e
and 
\begin{eqnarray} 
 \langle \Delta_{{\F} o}(\U_1,n)  \Delta^{*}_{{\F} o}(\U_2,m)  \rangle
 &=& 
\delta_{\U_1,\U_2} \, \left[ \frac{1}{L^2} P_{\F \F}(\U_1, (n-m)
  \Delta 
  z_c) \right. \nonumber  \\
&+& \left.  \frac{1}{N \, L^2} \sum_{\U} 
P_{\F \F}(\U,(n-m ) \Delta z_c)+
\delta_{n,m} \ \frac{\sigma^2_{\F \N}}{N} \right]
\label{eq:e10}
\end{eqnarray}
It is possible to simplify the sum over  $\U$ using Parseval's theorem
whereby 
\be 
\frac{1}{L^2}  \sum_{\U} P_{\F \F}(\U, ( n-m ) \Delta z_c) =
\xi_{\F}(( n-m) \Delta z_c) \,, 
\label{eq:e11}
\e
where $\xi_{\F}(\Delta z) =\langle \delta_{\F}(\t_a,z) 
\delta_{\F}(\t_a,z+ \Delta z)  \rangle $ is the one dimensional (1D)
correlation function of the fluctuations in the transmitted flux along
 individual quasar spectra. The 1D correlation $\xi_{\F}(\Delta z)$, or
 equivalently $\xi_{\F}(v_{\parallel})$,  is traditionally used to
 quantify the Ly-$\alpha$ forest along quasar spectra, and this has
 been quite extensively studied \citep{becker, coppolani, dodorico}.

Using $\bar{n}_Q=N/L^2$ to denote the quasar density on the sky, we
have 
\be 
 \langle \Delta_{{\F} o}(\U_1,n)  \Delta^{*}_{{\F} o}(\U_2,m)  \rangle
 = \delta_{\U_1,\U_2} \,  L^{-2} \, P_{\F \F o}(\U_1, n-m  )
\label{eq:e11a}
\e
where 
\be
P_{\F \F o}(\U,p) =  P_{\F \F}(\U, p  \Delta  z_c) 
+    \frac{1}{\bar{n}_Q} \left[ \xi_{\F}(p  \Delta z_c) +   
\delta_{p,0} \, \sigma^2_{\F \N} \right] 
\label{eq:e13}
\e Ideally, we would expect $P_{\F \F o}(\U,p)$ to provide an unbiased
estimate of $P_{\F}(\U,p \Delta z_c)$.  The discrete quasar sampling,
however, introduces an extra term $\xi_{\F}(p\Delta
z_c)/\bar{n}_Q$. The other term $\delta_{p,0} \sigma^2_{\F
\N}/\bar{n}_Q$ arises due to the pixel noise, and it contributes only
when we correlate a channel with itself.  Though assumed to be a
constant across the redshift range considered here, we note that
$\bar{n}_Q$ is infact a function of the magnitude limit of the survey
through the luminosity function $\frac{d \bar{n}_{Q}}{d m}(z)$
\citep{jiang} which accounts for the variability of quasar
luminosities. 

Radio interferometric observations directly measure
$\Delta_T(\U,z_n)$. We consider a
radio-interferometric array with  several antennas, each   of 
diameter $D$.  The antenna diameter and the field of view $L$
 are related   as $\lambda /D \approx L$, where
$\lambda$ is the observing wavelength. Each pair of antennas measures
$\Delta_T(\U,z_n)$  at a  particular $\U$ mode corresponding to
 $\U={\bf d}/\lambda$, where ${\bf d}$ is the antenna separation
 projected perpendicular to the line of sight. The 
baselines $\U$  corresponding to the different antenna pairs are, in
general,  arbitrarily distributed depending on the array
configuration.   The observed HI   fluctuation $\Delta_{To}(\U,n)$
at  two  different $\U$ values are correlated if $\mid \U_1-\U_2 \mid
\le  1/L$.  It is possible to combine the baselines where the signal
is correlated by binning  the $\U$ values  using cells of 
size $L^{-1} \times L^{-1}$. We then have the binned baselines at
$\U= (n_x \hat{i}+ n_y 
\hat{j} )/L$ ($n_x,n_y$ are integers)  which exactly match the
Fourier modes of the Ly-$\alpha$ 
signal.  The HI signal at different $\U$ values are now uncorrelated.
We then  have 
 \be 
{\Delta}_{{T} o}(\U,n) = \Delta_{T}(\U,z_n) +
\Delta_{T {\N}}(\U, n)  
\label{eq:e14}
\e 
where $\Delta_{T {\N}}$ is the corresponding noise contribution.  The 
noise in different channels and baselines is uncorrelated, and 
\be 
\langle \Delta_{{T} o}(\U_1,n) \Delta^*_{{T} o}(\U_2,m)
\rangle = \delta_{\U_1,\U_2} \, L^2\, P_{T T o}(U_1, n-m)
\label{eq:e14a}
\e
where 
\be
P_{T To}(U,p)= P_{TT}(\U, p   \Delta \,   z_c)  +   \delta_{p,0}
\N_T(\U)   \,.
\label{eq:e15}
\e
Here 
$\N_T(\U)$ is the noise power spectrum. For a single polarization
and a single baseline, this is given by 
\be
\N_T(\U)=\left(\frac{T^2_{\rm sys}\,}{2 \,  \Delta
  \nu_c \,   \Delta t} \right) \ \frac{[ \int d \Omega \,
    {\mathcal P}(\t)]^2}{[\int d   \Omega \, {\mathcal P}^2(\t) ]}
\label{eq:e16}
\e 
where  $T_{\rm sys}$ is the system temperature, 
$\Delta \nu_c$ the frequency  interval corresponding to $\Delta z_c$,
$\Delta t$ the integration time and ${\mathcal P}(\t)$ is the
normalised  power  
pattern of the individual antennas \citep{gmrt}. 
The exact value of the ratio of the two integrals in
eq. (\ref{eq:e16}) depend on the antenna design. It is convenient here
to express  eq. (\ref{eq:e16})  as 
\be
\N_T(\U)=\frac{T^2_{\rm sys}\,  L^2}{\chi  \, N_{pol} \,M(\U) \,  \Delta
  \nu_c \,   \Delta t} \,.
\label{eq:e17}
\e 
where  $N_{pol}$ is the number of polarizations being used, $M(\U)$ the
number of baselines in the particular cell corresponding to $\U$,  and
$\chi$ is a factor whose value depends on the antenna beam pattern
${\mathcal P}(\t)$. For the purpose of this paper it is reasonable to
assume a value $\chi=0.5$.

We have 
 \be
 \langle \frac{1}{2} \left[ {\Delta}_{{\F} o}(\U_1,n)
 {\Delta}^{*}_{{T} o}(\U_2,m) + 
   {\Delta}^{*}_{{\F} o}(\U_1,n) {\Delta}_{{T} o}(\U_2,m)
   \right] \rangle = \delta_{\U_1,\U_2} \, 
P_{\F T o}(\U_1,n - m )
 \label{eq:e18}
\e
where 
\be 
P_{\F T o}(\U,p) = P_{\F T}(\U, p \Delta z_c)
 \label{eq:e18a}
\e
which can serve as an estimator for the cross-correlation signal. 
It is, however, possible to increase the signal to noise ratio by
averaging over the entire redshift interval (or frequency band). We
therefore define  the estimator $ \hat{E}(\U,p)$ as 
\be
\E(p)=\frac{\sum_{n=1}^{N} \sum_{m=1}^{N} \frac{1}{2} 
\left[ \Delta_{{\F} o}(n) \Delta^{*}_{{T} o}(m) +
   \Delta^{*}_{{\F} o}(n) \Delta_{{T} o}(m)
   \right] \delta_{\mid n-m \mid,p}}{
\sum_{n=1}^{N} \sum_{m=1}^{N} \delta_{\mid n-m \mid,p}}
 \label{eq:e19}
\e
The various terms in eq. (\ref{eq:e19})   all refer to the
same $\U$ value which is not explicitly   shown for convenience of
notation. Note that we shall adopt this convention of not explicitly
showing $\U$ in several of the subsequent equations.

The estimator has the property that  
\be
\langle \E(\U,p) \rangle =  P_{\F T}(\U, p \Delta z_c) 
 \label{eq:e20} 
\e
{\it ie.} it is an unbiased estimator for the cross-correlation
signal. We next consider the covariance $\langle E(\U_1,p) \,
E(\U_2,q) \rangle - \langle E(\U_1,p) \rangle \, \langle
E(\U_2,q)\rangle $  which, we find, is zero  if $\U_1 \ \neq \U_2$.  We
therefore only consider 
\be
{\rm Cov}(p,q)=\langle \, \Delta E(p) \, \Delta E(q) \rangle = 
\langle E(p) \, E(q) \rangle - 
\langle E(p) \rangle  \, \langle E(q) \rangle
\label{eq:e21}
\e
where all the terms refer to the same $\U$. 

We note that  
\begin{eqnarray} 
&& \langle \left[ \Delta_{{\F} o}(n) \Delta^{*}_{{T} o}(m) +
   \Delta^{*}_{{\F} o}(n) \Delta_{{T} o}(m)
   \right] 
\left[ \Delta_{{\F} o}(s) \Delta^{*}_{{T} o}(t) +
   \Delta^{*}_{{\F} o}(s) \Delta_{{T} o}(t)
   \right] \rangle = \nonumber \\
&& 4 \, P_{\F T o}(n-m)  P_{\F T o}(s-t) 
+2 \,  P_{\F T o}(n-t)  P_{\F T o}( s-m) 
\nonumber \\
&+& 2 \, P_{\F \F o}(n-s)  P_{T T o}(t-m) 
\label{eq:e22}
\end{eqnarray}
whereby 
\be
{\rm Cov}(p,q)= \frac{\sum_{\alpha,\beta} \sum_{m,n,s,t}
P_{\alpha \beta o}(m-s)  P_{\alpha' \beta' o}( n-t)   
\ \delta_{\mid m-n \mid,p} \ \delta_{\mid s-t \mid,q}}{ 4 \
 \sum_{\alpha,\beta} \sum_{m,n,s,t}\
\delta_{\mid m-n \mid,p} \delta_{\mid s-t \mid,q}}
\label{eq:e23}
\e
where the variable $\alpha'$ has value $\F$ when $\alpha=T$ and
vice-verse, and $\beta,\beta'$ are defined in a similar way. 
We simplify eq. (\ref{eq:e23}) in two steps where we first have   
\be
{\rm Cov}(p,q)= \frac{ \sum_{\alpha,\beta}
\sum_{n=1}^{N_c-p}  \sum_{m=1}^{N_c-q}
P_{\alpha \beta o}(p+ m-n ) \  P_{\alpha' \beta' o}( m-n-q)
 }{8 (N_c-p)(N_c-q)}\,.
\label{eq:e24}
\e
Next, assuming that $ p \ge q$, we have 
\begin{eqnarray}
{\rm Cov}(p,q)&=& \sum_{\alpha,\beta} \frac{1}{8\, (N_c-q)}
\left\{\frac{1}{(N_c-p)} \sum_{m=1}^{N_c-p} (N_c-p-m)\left[ P_{\alpha
    \beta o}(p+ m ) P_{\alpha' \beta'o}(m-q) \right. \right.  \nonumber
  \\ &+& \left. \left.  P_{\alpha \beta o}(p-q+ m ) P_{\alpha'
    \beta'o}(m) \right] + \sum_{m=0}^{p-q} \left[ P_{\alpha \beta o}(p- m
  ) P_{\alpha' \beta' o}( q+m) \right. \right.  \nonumber \\ &+&
  \left. \left.  P_{\alpha \beta o}(p-q-m) P_{\alpha' \beta'o}(m) \right]
\right. \left.  \right\}
\label{eq:e25}
\end{eqnarray}
We use eq. (\ref{eq:e25}) to calculate the covariance ${\rm Cov}(\U,p,q)$ for
the individual $\U$ values. The expected signal is statistically
isotropic in $\U$, and  it is useful to bin the estimates  
of the power spectrum. We have the 'binned' estimator  
\be 
E_B(U_B,p) = \frac{\sum_i \, W_i \, E(\U_i,p)}{\sum_i \, W_i }
\label{eq:e26a}
\e
where $W_i$ refers to the weights assigned to the estimates at
different $\U_i$, and $U_B$ refers to the average baseline of the
particular bin 
\be 
U_B = \frac{\sum_i \, W_i \, \mid \U_i \mid}{\sum_i \, W_i } \,.
\label{eq:e26}
\e
The binned estimator has a covariance
\be 
{\rm Cov}_B(U_B,p,q)=\frac{\sum_i \, W^2_i \, cov(\U_i,p,q)}{N_{poin} \, \sum_{i,j} \,
  W_i W_j}\,.
\label{eq:e27}
\e
where the parameter $N_{poin}$ refers to the number of independent
pointings. This parameter  is introduced to  allow for the possibility
that we are combining estimates of the power spectrum from
observations in $N_{poin}$ independent  parts of the sky. It is quite clear
from eq. (\ref{eq:e27}) that this helps to increase the signal to
noise ratio as $1/\sqrt{N_{poin}}$. In subsequent parts of this paper, we
have used eq. (\ref{eq:e27}) to predict  the noise in  observations to
measure $P_{\F T}(U,\Delta z)$.

\section{Observational Considerations}

 The quasar redshift distribution peaks in the range  $z=2$ to $3$ 
\citep{sneider}, and   for our analysis we only consider the quasars 
in this redshift range.  For a quasar at a redshift $z_Q$, the region 
 $10,000 \, {\rm km \, s^{-1}}$ blue-wards of the quasar's Ly-$\alpha$
emission  is excluded from the Ly-$\alpha$ forest  due to the quasar's 
proximity effect.  Further, only the  pixels at least 1,000 $\rm km  \, 
s^{-1}$ red-ward of the quasar's Ly-$\beta$ and O-$\rm VI$  lines 
are considered to avoid the possibility of confusing  the Ly-$\alpha$ forest 
with the  Ly-$\beta$ forest  or the intrinsic O-$\rm VI$ absorption.  For a 
quasar at the fiducial redshift $z_Q=2.5$, the 
above considerations would allow the Ly-$\alpha$ forest to be measured
in the redshift range $ 1.96 \leq z \leq 2.39$ 
spanning an interval  $\Delta z =0.43$.  

We consider redshifted $21 \, {\rm cm}$  observations of bandwidth 
$ B = 128 \rm MHz$  covering  the frequency range
$ 355 \ \rm MHz$ to  $483 \rm MHz$ which corresponds to the 
redshift range $1.94\leq z \leq 3$ with bandwidth $B=1$ in redshift units.  
The Ly-$\alpha$ forest of any particular quasar will be measured 
in a smaller interval $\Delta z \approx 0.4$ which is the deciding
factor for the cross-correlation signal.  Thus, only a fraction 
(approximately $40 \%$) of the total number of quasars in this redshift
range $2$ to $3$ will contribute to the cross-correlation signal at 
any redshift $z$. We incorporate this in our  estimates by noting that 
$\bar{n}_Q$ in eq. (\ref{eq:e13}) refers to only $40 \%$ of all the quasars in 
the  entire $z$ range $2$ to $3$. 

Cosmic variance is  a limiting factor for measuring the BAO signal.
We first present a preliminary analysis to determine the observational
considerations so that the  first BAO peak is above the cosmic variance.  This depends
on $N_k$ the  number of independent Fourier modes in the  $k$ range 
$k_a= 0.03 \, \rm Mpc^{-1}$ to $k_b= 0.07 \, \rm Mpc^{-1}$  which 
covers the first BAO peak (Figure \ref{fig:baopower}). We have 
\be
N_k=\frac{V}{2 \pi^2} \int_{k_a}^{k_b} \,  k^2  \, dk
\label{eq:oc1a}
\e
where $V$ is the observational volume.  For observations covering the solid angle
$\Omega$ and  the $z$ range $1.94$ to  $3$, we have $N_k=2.38 \times 10^5 \, \Omega$.  
The BAO peak is a $10 \%$ feature in $P(k)$ and it is  necessary to 
divide  the interval  $k_a$  to $k_b$ into $N_{bin}$   bins in order to identify this 
feature.  An $N_{\sigma}$ detection  of the first BAO peak requires 
that the uncertainty in the observed power spectrum $\Delta P(k)/P(k) = \sqrt{N_{bin}/N_k}$ 
should be less than $1/(10 N_{\sigma})$.   We then have an estimate of the solid 
angle that needs to be observed for a $N_{\sigma}$ detection 
\be
\Omega = 1.89 \times 10^{-2} \left(\frac{N_{bin}}{5}\right)
\left(\frac{N_{\sigma}}{3}\right)^2  {\rm sr}\,.
\label{eq:oc2}
\e
We would like a single field of view of the radio-interferometer $\Omega=L^2$ 
to be large enough so as to cover this solid angle, whereby we need 
\be  
L = 8^{\circ}  \left(\frac{N_{bin}}{5}\right)^{0.5}
\left(\frac{N_{\sigma}}{3}\right)  .
\label{eq:oc3}
\e
A $3-\sigma$ detection with $N_{bin}=5$ requires a field of view  $L=8^{\circ}$
which can be achieved  if we have a radio-interferometric array where the individual
antennas are around $D=5 \, {\rm m}$ in diameter.  It is  advantageous to have 
a larger field of view, and we consider antennas of diameter $D=2 \, {\rm m}$ 
which gives a $L = 20^{\circ}$   field of view for which a $7.7 -\sigma$ detection is 
possible if $N_{bin}=5$.  Note that this sets the upper limit for the signal to noise
ratio (SNR)  that can be achieved in a single pointing.  In the next section
We present more detailed estimates of the SNR, taking into account various factors
like the discrete quasar sampling and the  noise in the quasar spectra and the radio
data. 

We next discuss the array  layout that would be required for these observations. 
Using $d/\lambda=U = k\, r/2\pi$, we estimate  that the Fourier modes $k_a$ and 
$k_b$ correspond to antenna separations $d_a = 15  $m and $d_b=36 $m respectively. 
These figures roughly set the range of antenna separations that would be required
in the radio-interferometric array.  Based on these considerations we  consider
a radio interferometric array which has  $N_{ann}$ antennas distributed such that all 
the baselines $\vec{d}$ within  $d_{max}=50 \, {\rm m}$ are uniformly sampled, 
whereby $M(\U)$ is independent of $\U$ and  we have $M(\U) \approx 4 \, (N_{ann}/100)^2$. 
Using this in eq. (\ref{eq:e17}), assuming $T_{sys}=100 \, {\rm K}$, $N_{pol}=2$, $\chi=0.5$ 
we have 
\be 
N_{T}=1.0 \times 10^{-3} \, [m {\rm K}]^2 \,     \left( \frac{100}{N_{ann}} \right)^{2} 
 \left( \frac{100 \, {\rm KHz}}{\Delta \nu} \right) 
 \left( \frac{1000 \, {\rm hrs}}{\Delta t} \right) \,.
\label{eq:oc1}
\e

\section{Detectability}
\begin{figure}
\psfrag{pc}[c][c][1][0]{{\bf\Large{ $P_{\F T}$}}}
\psfrag{l}[c][c][1][0]{{\bf\Large{ $\ell$}}}
\psfrag{a}[c][c][1][0]{{\bf\Large{ $(mK)$}}}

\begin{center}
 \mbox{\epsfig{file=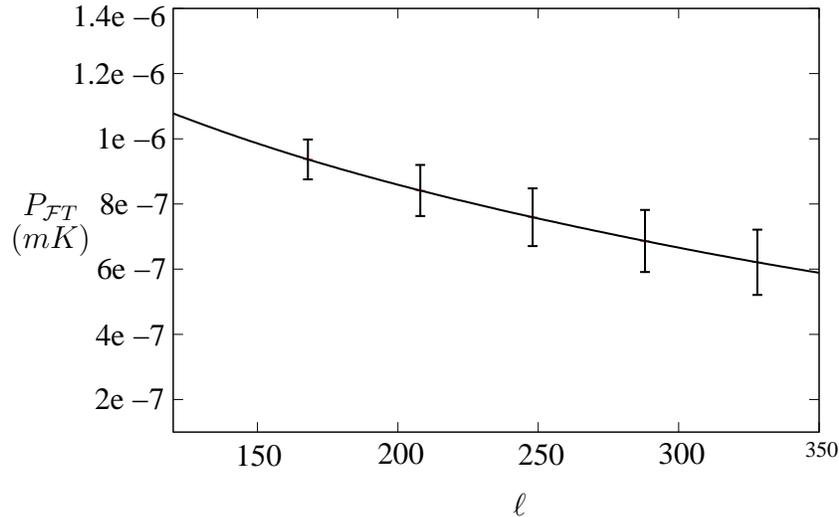,width=0.65505\textwidth,angle=0}}
\caption{The data points  shows  the binned transverse angular
  power spectrum  $P_{\F T}(\ell)$ for the cross-correlation signal at $z = 2.5$
 with $1-\sigma$ error bars. The error bars  correspond to the set of observational 
parameters (see text) that give a $5-\sigma$ detection of $P_{\F T}(\ell)$.}
\label{fig:clerror}
\end{center}
\end{figure}

\begin{figure}
\psfrag{NT}[c][c][1][0]{{\bf\Large{ $N_{T}$}}}
\psfrag{n}[c][c][1][0]{{\bf\Large{ $\bar{n}_Q \, (\rm deg^{-2})$}}}
\psfrag{a}[c][c][1][0]{{\bf\Large{ $(\rm mK^2)$}}}
\begin{center}
 \mbox{\epsfig{file=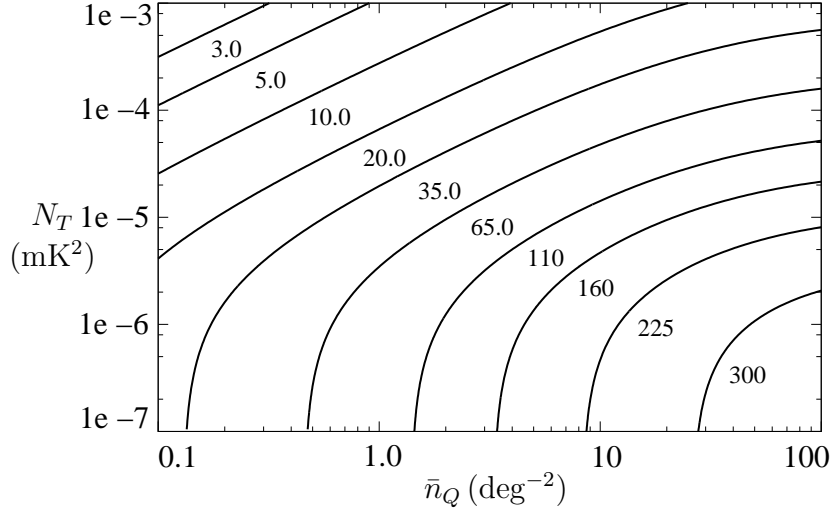,width=0.65505\textwidth,angle=0}}
\caption{Contours of ${\rm SNR}_1$, the signal to noise ratio for  
detecting the angular power spectrum in a single field of view.} 
\label{fig:contour}
\end{center}
\end{figure}

\begin{figure}
\psfrag{NT}[c][c][1][0]{{\bf\Large{ $N_{T} \, ( \rm mK^2)$}}}
\psfrag{l}[c][c][1][0]{{\bf\Large{ $\ell$}}}
\psfrag{SNR}[c][c][1][0]{{\bf\Large{ $ {\rm SNR}_{1}$}}}
\psfrag{deriv}[c][c][1][0]{{\bf\Large{\hspace{0.4cm}  $\mid \frac{d \,\ln {{\rm SNR}_1}
      }{ d \, \ln N_{T}}  \mid $}}}

\psfrag{pc}[c][c][1][0]{{\bf\Large{ $P_{\F T}$}}}
\begin{center}
 \mbox{\epsfig{file=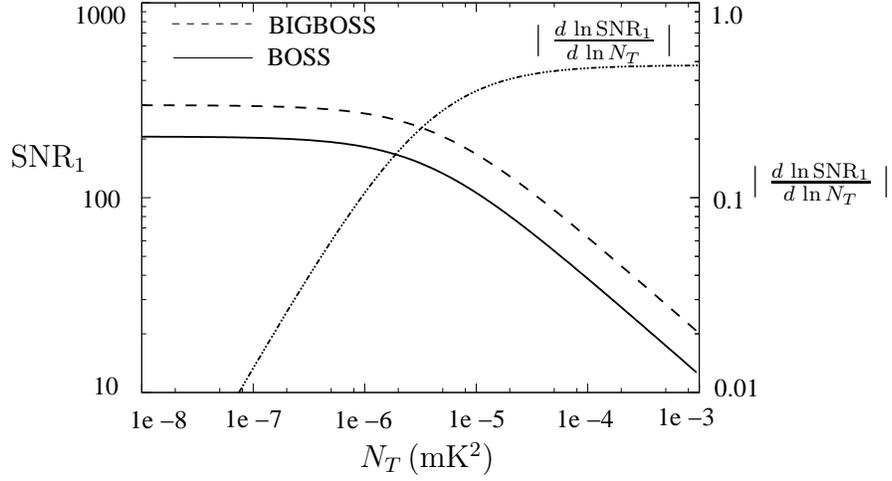,width=0.65505\textwidth,angle=0}}
\caption{The variation of ${\rm SNR}_1$ with $N_T$ for   $\bar{n}_Q=6.4 \, {\rm deg}^{-2}$
and $25.6 \, {\rm deg}^{-2}$ 
 corresponding to the BOSS and BIGBOSS respectively. We also show 
$\mid  \frac{d \,\ln {{\rm SNR}_1} }{ d \, \ln N_T} \mid $
which gives an estimate of how \snr1 scales  as $N_T$ is reduced. 
This is expected to have a value $0.5$ for large $N_T$ and drop  to 
$0$ as $N_T \rightarrow 0$.  We find that the ${\rm SNR}_1 \propto   {N_T}^{-0.5}$ 
scaling is approximately valid   for $N_T > 10^{-4} {\rm mK}^2$, and ${\rm SNR}_1$ falls 
slower than this for smaller values of $N_T$.}

\label{fig:snrvsnt}
\end{center}
\end{figure}

\begin{figure}
\psfrag{pc}[c][c][1][0]{{\bf{ $ \frac {{\Large P}_{\F  T}}{{\Large P}^{nw}_{\F  T}}$}}}
\psfrag{l}[c][c][1][0]{{\bf\Large{ $\ell$}}}
\psfrag{a}[c][c][1][0]{}
\begin{center}
 \mbox{\epsfig{file=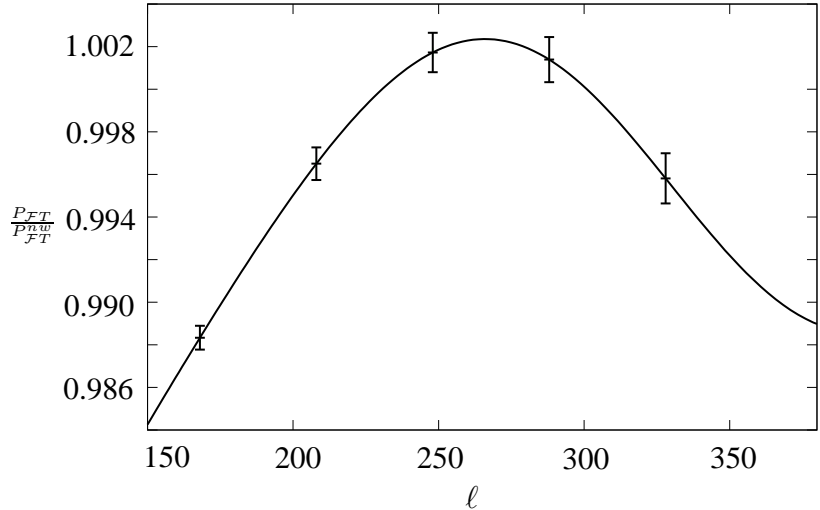,width=0.65505\textwidth,angle=0}}
\caption{This shows the first  BAO  peak in  the transverse angular
  power spectrum  $P_{\F T}(\ell)$ for the cross-correlation signal. To
  highlight the BAO we have divided $P_{\F T}(\ell)$ by  $P_{\F T}^{nw}(\ell)$
which has no-wiggles. The  binned data points and error bars  correspond to  $1-\sigma$ 
for  the set of observational 
parameters (see text) which  give a $5-\sigma$ detection of the transverse BAO.} 
\label{fig:clbaoerror}
\end{center}
\end{figure}

\begin{figure}
\psfrag{dkappa}[c][c][1][0]{{\bf\Large{ $\Delta \kappa_{\ell}(\Delta z)$}}}
\psfrag{dz}[c][c][1][0]{{\bf\Large{ $\Delta z$}}}
\psfrag{a}[c][c][1][0]{{\bf\Large{ $(mK)$}}}

\begin{center}
\mbox{\epsfig{file=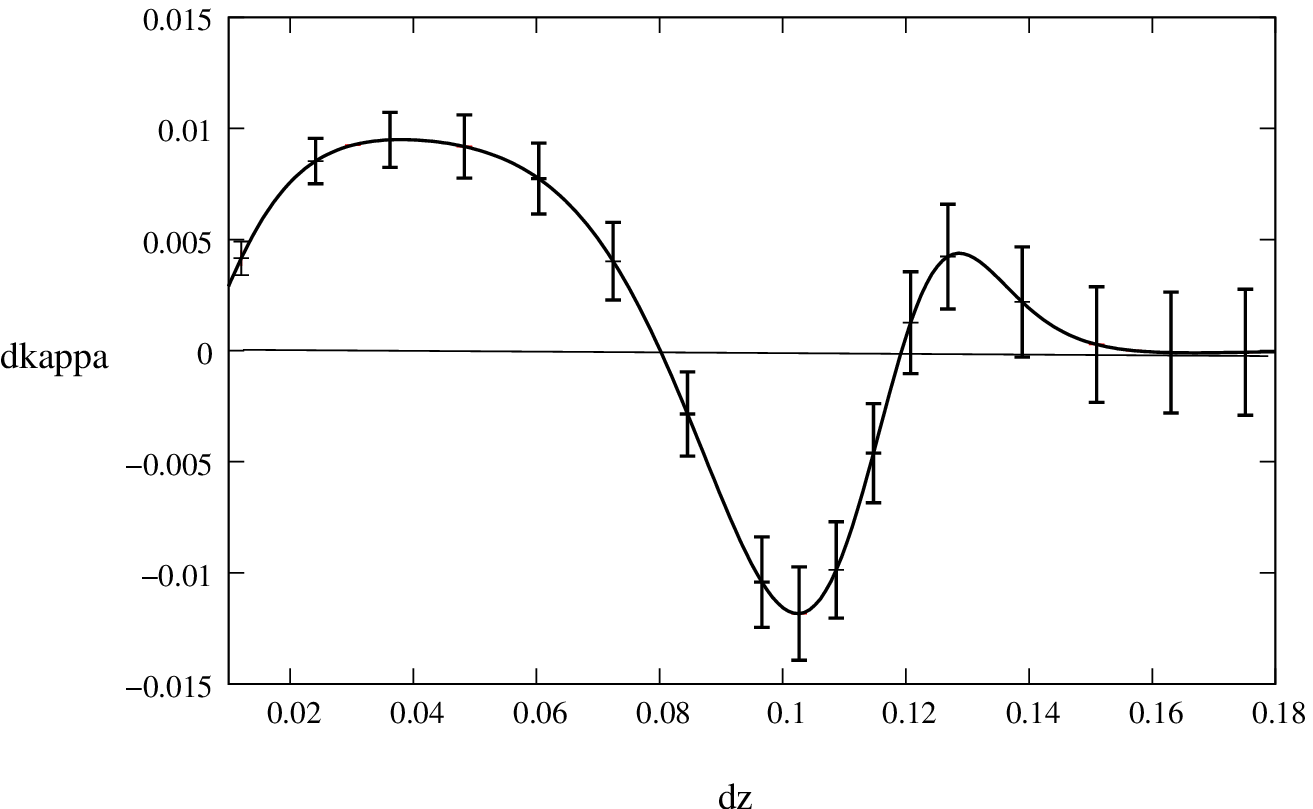,width=0.65505\textwidth,angle=0}}
\caption{This shows $\Delta \kappa_{\ell}(\Delta z)$, the difference between 
the BAO and no-wiggle models   for the
  cross-correlation signal at the $\ell$ bin centered at $\ell =  230$.
The  binned data points and error bars  correspond to  $1-\sigma$ 
for  the set of observational 
parameters (see text) which  give a $5-\sigma$ detection of the radial  BAO.} 
 
\label{fig:deltakappa}
\end{center}
\end{figure}
In the previous section we have discussed several observational
considerations which essentially determine the redshift interval and
angular scales that need to be covered in order to detect the imprint
of the BAO in the cross-correlation signal .  We have seen that
observations covering the redshift interval from $z=2$ to $3$ using a
radio interferometric array with antennas of diameter $2 \, {\rm m}$
each are well suited for this purpose.  Further, for the purpose this
paper, we have made the simplifying assumption that the array layout
is such that it uniformly samples all the baselines within $50 \, {\rm
m}$.  Given this observational framework, we now estimate the required
observational sensitivity for a detection of the BAO signal.

The variance of the cross-correlation estimator ${\rm
Cov}(p,p)=\langle \, \Delta E(p) \, \Delta E(p) \, \rangle$ is a sum of
several terms (eq. \ref{eq:e25}) of the form $P_{\F T}(p+m) \, P_{\F
T}(p-m)$ and $P_{T T o}(p+m) \, P_{\F \F o}(p-m)$, where $P_{\F T}$
refers to the cross-correlation signal that we are trying to detect.
The terms $P_{\F \F o}$ and $P_{T T o}$ refer to the auto-correlation
of the Ly-$\alpha$ and redshifted 21-cm observations respectively.
Note that $P_{\F \F o}$ and $P_{T T o}$ both have contributions from
observational noise, and hence they differ from the respective
cosmological auto-correlations signals $P_{\F \F}$ and $P_{T T }$.

We first consider $P_{\F \F o}$ which refers to the Ly-$\alpha$
forest.  In addition to $P_{\F \F}$, this also has a contribution from
noise that arises due to the discrete QSO sampling.  This noise, which
arises from the discrete sampling of quasars, is proportional to
$\bar{n}_Q^{-1}$ (eq. \ref{eq:e13}).  To recapitulate, $\bar{n}_Q$
refers to approximately $0.4$ times the total angular density of
quasars in the redshift range $2$ to $3$.  This noise also depends on
$\xi_{\F}$ for which we have used the values from
\cite{becker}. Further, we have assumed that the Lyman-$\alpha$ forest
spectra are all measured with a high sensitivity such that the pixel
noise contribution has a value $\sigma_{\F N}^2= 0.04$.

We next consider $P_{T T o}$ which refers to the redshifted 21-cm
observations.  In addition to the cosmological signal $P_{T T }$, this
also has a contribution $N_T$ from the system noise
(eq. \ref{eq:e15}).  The quantity $N_T$ refers to the noise power
spectrum of the radio interferometric observations.  The value of
$N_T$ depends on several observational parameters.  In the present
context, for observations with a fixed channel width $\Delta \nu$,
$N_T$ depends only on the number of antenna $N_{ann}$ and the
observing time $\Delta t$ (eq. \ref{eq:oc1}).  We see that there are
many different combinations of $N_{ann}$ and $\Delta t$ which will
give the same value of $N_T$.

In the present analysis we have used $\bar{n}_Q$ and $N_T$ to
 parametrize the respective sensitivities of the QSO survey and the
 redshifted 21-cm observations We have $P_{\F \F o} \rightarrow P_{\F
 \F}$ and $P_{T T o} \rightarrow P_{T T}$ in the limits $\bar{n}_Q
 \rightarrow \infty$ and $N_T \rightarrow 0$ respectively.  The limit
 $\bar{n}_Q \rightarrow \infty$ and $N_T \rightarrow 0$ sets the upper
 bound for the SNR, which also corresponds to the cosmic variance. The
 additional noise contributions arising from $\bar{n}_Q$ and $N_T$
  will degrade the SNR to a value which is lower than the cosmic
 variance.  In this work we would like to determine the range of
 $\bar{n}_Q$ and $N_T$ where it will be possible to detect the BAO
 feature.  The entire subsequent discussion refers to the fiducial
 redshift $z=2.5$.  Though we expect all the quantities to vary across
 the redshift range $z=2$ to $z=3$, we anticipate that the values at
 $z=2.5$ will be representative of the entire $z$ range.

We first consider the transverse angular power spectrum $P_{\F
T}(\ell) \equiv P_{\F T} (U, \Delta z)$ at $\Delta z =0$.  It is also
possible to study the transverse angular power spectrum by considering
the $U$ dependence of $ P_{\F T} (U, \Delta z)$ while holding $\Delta
z$ fixed at some other value $\Delta z>0$. The signal, however, is
maximum for $\Delta z =0$.  Further, it is also possible to increase
the SNR by suitably combining the estimates at different values of
$\Delta z$. However, we do not consider these possibilities here, and
we only consider the transverse angular power spectrum at $\Delta
z=0$.  The $\ell$ range $125 \leq \ell \leq 330$, we have seen, is
adequate for detecting the first BAO peak (Figure
\ref{fig:baoangular}).  We assume that the baselines $\U$ within this
$\ell = 2 \pi \mid \U \mid$ range have been divided into $5$ bins of
equal $\Delta U$, and we consider the results (eg. Figure
\ref{fig:clerror}) at the average $\ell$ values (eq. \ref{eq:e26})
corresponding to each of these bins. We have set $W_i=1$ whereby all
all the baselines within a bin contribute equally to the binned
estimator (eq. \ref{eq:e26a}). Here we have chosen a channel width of
$100 \rm KHz$ for the 21-cm data, and we assume that the Ly-$\alpha$
forest is smoothed at the corresponding velocity width of $\approx 100
\rm Kms^{-1}$.  The cross-correlations signal starts to decorrelate
(Figure \ref{fig:kappa1}) if the frequency channel width is larger. We
have used the results for the central $\ell$ bin, which roughly
corresponds to the center of the first BAO peak, to assess the overall
SNR.  Note that in the subsequent discussion we use\snr1 to refer to
the signal to noise ratio for a single field of view, and SNR (or
total SNR) to refer to the general situations where there are
$N_{poin}$ ($\ge 1$) independent pointings of the $20^\circ \times
20^\circ $ field of view.  Figure \ref{fig:contour} shows \snr1
contours as a function of $\bar{n}_Q$ and $N_T$.  We see that the
\snr1 does not increase very significantly beyond $\bar{n}_{Q} > 60
\rm deg^{-2}$ or $N_T < 10^{-7} \rm mK^2$.  We have the maximum value
of the \snr1 at the bottom right corner of Figure \ref{fig:contour}
where both of these conditions are satisfied.  This corresponds to the
$\bar{n}_Q \rightarrow \infty$ and $N_T \rightarrow 0$ limit, and the
maximum \snr1, which corresponds to the cosmic variance.  has a value
\snr1=$300$.  Note that the \snr1 contours in Figure \ref{fig:contour}
refer to the detection of the cross-correlation signal $P_{\F
T}(\ell)$, and not the BAO which is just a $1 \%$ feature in $P_{\F
T}(\ell)$.  We find that a $5-\sigma$ detection of $P_{\F T}(\ell)$ is
possible for $\bar{n}_{Q} \sim 0.1 \, \rm deg^{-2}$ and $N_{T} \sim
10^{-4} \rm mK^2$.  The required QSO density is well within the scope
of present observational capabilities, for example the currently
available SDSS \citep{sneider} has a total QSO number density of $\sim
1 \ \rm deg^{-2}$ implying $\bar{n}_{Q} \sim 0.4 \ \rm deg^{-2}$ which
is in excess of the required QSO number density.  The requirement that
$N_{T} \sim 10^{-4} \rm mK^2$.  can be achieved in $1,000 \, {\rm
hrs}$ if we have $300$ antennas.  A longer observation will be
required if we have fewer antennas (eq: \ref{eq:oc1}).  Figure
\ref{fig:baoangular} shows the binned data points and the $1-\sigma$
error-bars that are expected in such an observation.

The first BAO peak is a $1\, \%$ feature in $P_{\F T}(\ell)$, and a
$5-\sigma$ detection of the BAO peak requires an SNR of $500$ for
$P_{\F T}(\ell)$. It is not possible to reach an SNR greater than
$300$ with a single field of view (Figure \ref{fig:contour}), and it
is necessary to consider multiple pointings.  The total signal to
noise ratio increases as \be {\rm SNR}=\sqrt{N_{poin}} \, {\rm SNR}_1
\, ,
\label{eq:det}
\e 
when   the  number of independent fields of view is  increased.
We require  $100$ independent fields fields of view   to detect the BAO 
with  the currently available SDSS.   This  exceeds  the angular area of the 
SDSS,  and is not  viable. 
It is necessary to consider a survey with a higher quasar density. 
The upcoming  BOSS \citep{mcd1}   is expected to have a QSO density of 
$16 \ \rm deg^{-2}$ which corresponds to  $\bar{n}_{Q} = 6.4 \ \rm deg^{-2}$. 
The BOSS \footnote{http://cosmology.lbl.gov/BOSS/} survey is expected to 
cover  $\sim 10,000 \rm deg^2$  of the sky, and we could ideally have 
$N_{poin}= 25 $ independent  pointings  of the   $20^\circ \times 20^\circ $ field of view. 
We see that, with BOSS,  it is possible to achieve an SNR larger than $100$ in a single 
field of view (Figure \ref{fig:contour}).  Therefore, 
 a $5-\sigma$ detection of the first BAO
peak is possible with  $N_{poin} \le 25$,  which is within the total angular
coverage of BOSS.    Figure \ref{fig:snrvsnt} shows how \snr1 varies with $N_T$
for a single field of view of BOSS.  
We see that the \snr1 scales as $N_T^{-1/2}$ for  large $N_T$ 
where the system noise in the radio observations is much larger than  the HI signal. 
 In this regime we have ${\rm SNR} \propto \sqrt{N_{poin}/N_T}$ 
  {\it ie.}       relatively 
shallow observations of a large number of fields of view, or deep observations   
of a few fields of view would both give the same  SNR provided that the total observing 
time   $N_{poin} \, \Delta t$   (eq. \ref{eq:oc1}) is the same in  both of  these 
situations. 
 Ideally, it is most advantageous to work in this  region. However,  we see that 
the  ${\rm SNR} \propto N_T^{-1/2}$scaling only holds at $N_T > 10^{-4} \, {\rm mK}^2$   
where the \snr1 is pretty low ($\sim 40$), and a $5-\sigma$ detection of the BAO peak
is not possible within the total angular coverage of BOSS. It is  necessary 
to consider observations that go deeper than $10^{-4} \, {\rm mK}^2$.     
The increases in  \snr1 is slower than $N_T^{-1/2}$  for 
$N_T \le 10^{-4} \, {\rm mK}^2$, and the \snr1 saturates at
 $N_T \le 10^{-7} \, {\rm mK}^2$  where $N_T$ is much smaller  than the HI signal. 
The range 
$ 10^{-5} \, {\rm mK}^2 \ge  N_T \ge 10^{-6} \, {\rm mK}^2$, where the\snr1 is in 
the range   $100$ to $200$, is relevant  for detecting the BAO  with the 
cross-correlation signal using BOSS.   In this region it  is optimal  to increase
 the total  SNR by increasing the number of fields of view instead of increasing 
the depth of the  individual observations. We find that we have ${\rm SNR}_1=100$ for 
 $ N_T=1.1 \times 10^{-5} \, {\rm mK}^2$,  and a $5-\sigma$ detection 
of the first BAO peak is possible with $N_{poin}=25$ fields of view. 
Alternatively, a $5.8-\sigma$ detection is possible with $N_{poin}=16$ and
 $N_T=6.25 \times  10^{-6} \, {\rm mK}^2$. The scaling is approximately 
 ${\rm SNR} \propto N_{poin}^{0.5} \,  N_{T}^{-0.38}$ here, and the total observing
time is smaller if we consider  the shallower observations. 
Figure \ref{fig:clbaoerror}  shows the expected binned data points and  error-bars 
for the detection of the first BAO peak. 

The  BIGBOSS \citep{bigboss} has been conceived as the successor to the upcoming BOSS
QSO survey.  BIGBOSS may achieve  a QSO density of $\sim 64 \, \rm deg^{-2}$
which corresponds to $\bar{n}_Q=25.6 \ {\rm deg}^{-2}$.  This is very close to the
 region where we have the cosmic variance limit, 
and the limiting \snr1 of $\sim 300$ can be achieved if 
 $N_T \approx 2 \times 10^{-6} \, {\rm mK}^2$ (Figure \ref{fig:contour}). 
For this value of $N_T$, 
a $5-\sigma$ detection of the first BAO peak is possible with only two fields 
of view. However this is not the optimal observational strategy.  
For BIGBOSS, the 
behaviour of the \snr1 as a function of $N_T$ is very similar to that for BOSS, except that
the \snr1 values are $1.5$ times larger  (Figure \ref{fig:snrvsnt}).
The range 
$3.3 \times  10^{-5} \, {\rm mK}^2 \ge  N_T \ge 10^{-6} \, {\rm mK}^2$, where the\snr1 is in 
the range   $100$ to $300$, is relevant  for detecting the BAO  with the 
cross-correlation signal using BIGBOSS.
It is most advantageous to consider relatively  shallow  observations with 
 $N_T=3.3 \times  10^{-5} \, {\rm mK}^2$,   for which a detection is possible with 
 $N_{poin}=25$. The scaling here is 
${\rm SNR} \propto N_{poin}^{0.5} \,  N_{T}^{-0.43}$, which is quite close  to the
 $N_T^{-0.5}$  behaviour. 

We next consider the detection of the radial oscillations. 
We have seen  (Figure \ref{fig:baokappa}) that the
 radial oscillations occur in the $\Delta z$ range  $0.04$ to $0.14$. 
where, in the absence of the BAO,  the cross-correlation signal is anti-correlated 
{\it ie.} $\kappa_{\ell}(\Delta z) \le 0$, and relatively weak with 
$\mid \kappa_{\ell}(\Delta z) \mid \sim 0.01$ . The BAO introduces a ringing feature 
which is highlighted in Figure \ref{fig:deltakappa} which shows 
$\Delta \kappa_{\ell}(\Delta z)$ which is the difference between the BAO 
and the the no-wiggles model. We see that the  deviation  could be as large as 
$\Delta \kappa_{\ell}(\Delta z) \approx 0.01$.
Therefore,  the net effect of the radial oscillations is a 
 $\sim 1 \%$ deviation relative to $P_{\F T}(\ell)$, which is comparable to 
the deviation introduced by the angular oscillations. Though the angular and the 
radial oscillations both introduce  $\sim 1\%$ deviations relative to the no-wiggles
model, we do not  expect that every  observation
which is  capable of detecting the angular oscillations will also  be able to 
detect the radial oscillations.  The radial oscillations occur at $\Delta z \approx 0.1$,
and we have only $(B/ \Delta z) \sim 10$ independent estimates in our observation. 
In contrast, the tangential feature occurs at $\theta_s \approx 1.38^{\circ}$ and we 
have $(20^{\circ}/20^{\circ})^2 \approx 210$  independent estimates. We thus expect  the
radial oscillations to have a lower SNR  in comparison to the tangential oscillations
discussed earlier.
The BAO signal is maximum in the vicinity of $\ell \approx 250$,  and we have collapsed 
 the three central $\ell$ (or $U$) bins in order to enhance the SNR 
 for the radial oscillations.  We find that this enhances the SNR by a factor 
of approximately $\sqrt{2.7}$. 
The frequency channel width is maintained at  
 $\Delta \nu=100 \ {\rm KHz}$ which is the  same as the value used for the 
tangential oscillations. Note that the errors in the 
$P_{\F T }(U,\Delta z)$ values estimated at  different $\Delta z$ will, in general, 
be correlated and a rigorous error analysis would require us to calculate the full
covariance matrix (eq. \ref{eq:e25}). We have not used the 
covariance matrix in the present  analysis, 
 Instead, we have assessed the SNR for a detection of the radial oscillations by using 
just  the  single  value at  $\Delta z=0.1$  where the deviation from the no-wiggles 
model is maximum (Figure \ref{fig:deltakappa}).  
 We find that a $5 - \sigma$ detection is possible with BOSS if  we  observe $25$ fields of
 view  with   $N_T=  6.25 \times 10^{-6} \, {\rm mK}^2$.  We require 
$N_T=1.7 \times 10^{-5} \, {\rm mK}^2$ for a similar detection with BIGBOSS. 

\section{Summary and Discussion}
In this paper we have developed a theoretical formalism for  estimating the 
cross-correlation signal between the fluctuations in the Ly-$\alpha$ forest and the 
fluctuations in the redshifted 21-cm emission from neutral hydrogen. Both of these
quantities are measured as functions of  frequency (redshift) and  angular scale. 
Consequently, we have  used the Multi-frequency Angular Power Spectrum (MAPS) 
to quantify the statistical properties of the cross-correlation signal. This 
deals directly with the observed quantities, and  retains the distinction between 
the angular and the frequency information. This, as we shall elaborate shortly, is 
very important in the light of foreground removal and continuum subtraction. 

Continuum fitting and subtraction is a very critical step in calculating 
$\delta_{\F}$ for the Ly-$\alpha$  forest  and several different methods have been 
proposed  for handling this  \citep{pspec4, mcd06}.  Errors in continuum subtraction
can be a serious problem for  the Ly-$\alpha$  forest auto-correlation signal
 \citep{kim04}. The redshifted 21-cm signal is buried deep under astrophysical
foregrounds which are several orders of magnitude larger
\citep{shaver99,fg4,fg1,fg6, fg3, pen09, fg10}.  Several different
techniques have been proposed for separating the cosmological 21-cm
signal from the foregrounds  \citep{ fg3, jelic, fg7, fg8, fg10}.
Foreground removal is a rather severe  problem for observations
of the redshifted 21-cm auto-correlation signal.  In addition to the 21-cm 
signal, the foregrounds make a very large contribution to the expectation 
value of the auto-correlation estimator. However, 
 the foregrounds are believed to have a slowly varying, smooth
frequency  (or $\Delta z$) dependence which is  quite distinct from the 
signal which decorrelates rapidly with increasing $\Delta z$ 
(Figure \ref{fig:kappa1}).  It is therefore possible to remove  the foregrounds 
from the measured MAPS ($P_{\F T o}(\ell,\Delta z)$ ) 
 by subtracting  out any component that varies slowly with $\Delta z$. 
In fact, \citet{fg11} have recently  used MAPS to analyze 610-MHZ GMRT  observations and
 show that it is possible to remove the foregrounds from  the m auto-correlation
by fitting and subtracting out slowly varying polynomials in $\Delta \nu$. 

We do not expect the continuum  in the Ly-$\alpha$ forest to have any correlation
with the foregrounds in the redshifted 21-cm observations,  and  consequently
they will  not contribute to the expected cross-correlation signal.
The continuum and the foregrounds will, however, appear as extra contributions
to the variance of the cross-correlation estimator. 
Since these contributions appear in the variance, it is possible to  
reduce these by  combining different independent estimates
of the cross-correlation.  We can  reduce the continuum and foreground 
contributions  by combining  estimates
of the cross-correlation at different baselines $\U$ and different fields of 
view. The problem, therefore, is much less severe in comparison  to the
 auto-correlation. 
Further, the continuum and the foregrounds are both expected  to have a slowly
 varying frequency  (or $\Delta z$) dependence and it should be possible to remove
these from the measured MAPS by  subtracting  out the  component that varies slowly with
 $\Delta z$.  We plan to perform a detailed analysis of these issues in future. 

Our study shows that it is possible to have a  $5-\sigma$ detection of the imprint of the 
first BAO peak in the cross-correlation signal   using 
BOSS, an upcoming QSO survey. For this, we have considered a radio interferometric array 
that covers the $z$ range $z=2$ to $3$ using antennas of size $2 \, {\rm m}
\times 2 \, {\rm m}$ which have a $20^{\circ} \times  20^{\circ}$ field of view. 
We find that  we need to observe $25$ fields of view, approximately
the full angular coverage of BOSS.    
with a noise level of 
$N_T=1.1 \times 10^{-5} \, {\rm mK}^2$  and $N_T=  6.25 \times 10^{-6} \, {\rm mK}^2$ 
in order to achieve  a 
$5-\sigma$ detection of the angular  and radial oscillations respectively. 
The corresponding noise levels are 
$N_T=3.3 \times 10^{-5} \, {\rm mK}^2$  and $N_T=  1.7 \times 10^{-5} \, {\rm mK}^2$ 
for BIGBOSS whose quasar density is expected to be four times larger than that of 
BOSS. 

We now briefly discuss how it may be possible 
to carry out such observations. In our analysis 
we have made the simplifying assumption that the antennas are distributed
such that all the baselines within $50 \, {\rm m}$ are uniformly sampled. 
 We consider an interferometric array with $N_{ann}=400$ 
antennas which  roughly corresponds to the maximum number of $2 \, {\rm m} \times 
2 \, {\rm m}$ antennas that can fit in a $50 \, {\rm m} \times 50 \, {\rm m}$ 
region.  We see that we need $\Delta t = 5700 \, {\rm hr}$ and 
$10,000 \, {\rm hr}$ of observation per field
of view (\ref{eq:oc1})  to reach  the  noise levels  required to detect
the angular and radial oscillations respectively with BOSS. 
The corresponding figures are  $1900 \, {\rm hr}$ and $3700 \, {\rm hr}$ 
for the BIGBOSS.  Note that we it is required to 
observe a single field of view for $8 \, {\rm hr}$ a day for 
a whole year in order to achieve $3000 \, {hrs}$ of  observing time.  

It is quite evident that we require to observe $25$ fields of view, with  
$2$ to $3$ years of dedicated observations for each field,  
in  in order to detect the BAO. Operating sequentially, considering 
one field after the next, the required observations would possibly run over 
a period of $50 \, {\rm yr}$ to a century, which raises the need to consider 
alternative observational strategies.  
 It is a viable  possibility  to  have antennas that  can 
simultaneously observe several independent  fields of view.  However, it is unlikely
(if not impossible) to have antennas that can simultaneously observe $16$ or $25$  such
$20^{\circ} \times 20^{\circ}$  fields of view.   For the purpose of this discussion, 
we assume that we have antennas that can simultaneously observe $4$ fields of view. 
We then see that it would approximately require observations over a decade (or more) 
in order to detect the BAO. Another possibility is to have 
 several radio interferometric arrays, each located at a different  location
 and observing a different parts of the sky. Four to five separate arrays, each
 with $400$  antennas, 
would be required to carry out these observations in the span of a few years. 
 It is important to note 
that it may be possible to reduce the observational requirements to some
 extent by optimally distributing the baselines instead of considering them to be
uniformly distributed. We propose to investigate  these issues   in a future study.

Observations of the BAO can be used to constrain the values of various cosmological
parameters. The equation of state of the Dark Energy is particularly important
in this context. In this paper we have mainly  estimated the range of observational
parameters for which  it  will be possible to detect the  BAO using  the 
cross-correlation signal. We plan to study a variety of issues including
the optimal array configuration  and cosmological
parameter estimation in future.

\section{Acknowledgement}
TGS would like to acknowledge Centre For Theoretical Studies, IIT
Kharagpur for using its various facilities. TGS would also like to
thank Tirthankar Roy Choudhury for useful discussions and help.

\end{document}